\documentclass[aps,showpacs,floatfix,twocolumn,amsmath,amssymb]{revtex4}
\usepackage{epsfig}  

\begin{document}

\title{N to $\Delta$ transition amplitudes from QCD sum rules}
\author{Lai Wang and Frank X. Lee}
\affiliation{Physics Department, The George Washington University,
Washington, DC 20052, USA}
 
\begin{abstract}
We present a calculation of the N to $\Delta$ electromagnetic transition amplitudes
using the method of QCD sum rules. A complete set of QCD sum rules is derived for the
entire family of transitions from the baryon octet to decuplet.
They are analyzed in conjunction with the corresponding mass sum rules
using a Monte-Carlo-based analysis procedure.  The performance of each of the sum rules
is examined using the criteria of OPE
convergence and ground-state dominance, along with the role of the transitions
in intermediate states. Individual contributions from the u, d and s quarks
are isolated and their implications in the underlying dynamics are explored.
Valid sum rules are identified and their predictions are obtained.
The results are compared with experiment and other calculations.

\end{abstract}
\vspace{1cm}  
\maketitle

\section{Introduction}
\label{intro}

The determination of low-energy electromagnetic properties of
baryons, such as charge radii, magnetic and quadrupole
moments, has long been a subject of interest in nucleon structure studies. 
For example, the E2/M1 ratio of the transition amplitudes 
in the process $\gamma N \rightarrow \Delta$ is a signature of the 
deviation of the nucleon from spherical symmetry.

The first serious attempt to determine the ratio from quantum chromodynamics (QCD), 
the fundamental theory of the strong interaction, was the lattice 
calculation of Leinweber {\it et al.}~\cite{Derek93}. 
The calculation has been updated recently by Alexandrou {\it et al.}~\cite{alex04,alex08} 
with better technology and statistics.
In this work, we present a calculation of the transition amplitudes using the SVZ 
method of QCD sum rules~\cite{SVZ79} which is another nonperturbative approach
firmly-entrenched in QCD with minimal modeling. 
The approach provides a general way of linking hadron phenomenology 
with the interactions of quarks and
gluons via only a few parameters: the QCD vacuum condensates and
susceptibilities. The analytic nature of the approach offers an unique perspective 
on how the properties of hadrons arise from nonperturbative interactions
in the QCD vacuum, and gives a complimentary view of the same physics 
to the numerical approach of lattice QCD.
It has been successfully applied to almost every aspect of strong-interaction physics,
including the only other attempt on the N to $\Delta$ transition that we know of 
by Ioffe~\cite{Ioffe84} over 25 years ago.
It was a limited study conducted in the context of the magnetic moments of the nucleon, 
and the results were inconclusive due to large contamination 
from the excited-state contributions.

The above discussion only concerns with the classic QCD sum rule method by SVZ.
There are a number of studies on N to $\Delta$ transitions in the so-called light cone 
QCD sum rule method (or LCSR)~\cite{Rohrwild07,Aliev06}. 
The main difference is that it employs as input light-cone wavefunctions, 
instead of vacuum condensates.  
One advantage of the light-cone QCD method is that one can
compute the transition form factor at non-zero momentum transfer
relatively easily~\cite{Belyaev96,Braun06}.

Our goal is to carry out a comprehensive study of the N to $\Delta$ transition
in the SVZ QCD sum rule method with a number of features.
First, we employ generalized interpolating fields which allow us to
use the optimal mixing of interpolating fields to achieve the best
match in the QCD sum rules. The interpolating field used previously is a special case.
Second, we derive the complete tensor structure and construct a new set of QCD sum rules.
The previous study was quite limited in its tensor structure.
Third, we study all tranisitons from the octet to te deuplet, not just 
the proton channel.
Fourth, we perform a Monte-Carlo uncertainty analysis which has become the standard
for evaluating errors in the QCD sum rule approach.
The advantage of such an analysis is explained later.
Fifth, we use a different procedure to extract the transition amplitudes
and to treat the transition terms in the intermediate states. Our
results show that these transitions cannot be simply ignored. Sixth,
we isolate the individual quark contributions to the transition amplitudes
and discuss their implications in the underlying quark-gluon dynamics.
A similar study has been performed for the magnetic moments of octet baryons~\cite{Wang08}.

The paper is organized as follows. In Section~\ref{phys},
the physics of N to $\Delta$ transition is reviewed. 
In Section~\ref{meth}, the external field method of QCD sum rules is applied to 
this process. Master formulas are calculated using the full sets of 
interpolating fields for the octet and decuplet baryons.
Then the full phenomenological representation and QCD side are calculated,
leading to a complete set of QCD sum rules for the entire family of transitions 
from the octet to the decuplet. Section~\ref{ana} describes our numerical procedure
to extract the transition amplitudes.
Section~\ref{res} summarizes the results and gives a comparison of our results 
with experiment and other calculations.
The conclusions are given in Section~\ref{con}.

\section{Physics of N to $\Delta$ transitions}
\label{phys}

\subsection{Definition of the form factors}
The concept of form factors plays an important role in
the study of internal structure of composite particles. The
non-trivial dependence of form factors on the momentum transfer
({\it i.e.}, its deviation from the constant behavior) is 
a signal of the non-elementary nature of the investigated particle.
The current matrix element for the $\gamma N \to\Delta$ transition can be written as 
\begin {equation}
 <\Delta (k',s')|j_\mu |N(k,s )> =\bar u^\rho (k',s')\Gamma_{\rho \mu} u(k,s),
\end {equation}
where $k,s$ and $k',s'$ denote initial and final momenta and spins,
$j_\mu$ the electromagnetic current,
$u$ the spin-1/2 nucleon spinor, and $u^\rho$ the Rarita-Schwinger spinor 
for the spin-3/2 $\Delta$.
This matrix element is the most general form required for describing on-shell 
nucleon and $\Delta$ states with both real and virtual photons.
There exist different definitions in the literature for the form factors 
describing the vertex function $\Gamma_{\rho \mu}$.
One is in terms of $G_1$, $G_2$, and $G_3$ by Jones-Scadron~\cite{JS73}:
\begin {equation} 
\Gamma _{\rho \mu} = 
G_1 (q^2 ){\rm H}_{\rho \mu}^1 + G_2 (q^2 ){\rm H}_{\rho \mu}^2 + 
G_3 (q^2 ){\rm H}_{\rho \mu}^3,
\label{js1}
\end {equation}
where the invariant structures are
\begin{equation}
\begin{array}{rcl}
 H_{\rho \mu}^1  &=&  q_\rho  \gamma _\mu   - q \cdot\gamma g_{\rho \mu};  \\
 H_{\rho \mu}^2  &=&  q_\rho P_\mu - q\cdot P g_{\rho \mu};  \\
 H_{\rho \mu}^3  &=& q_\rho q_\mu   - q^2 g_{\rho \mu}.
 \end{array}
\label{js2}
\end{equation}
Here $q=k'-k$ is the four-momentum transfer. 
For real photons ($q^2=0$), only $G_1(0)$ and $G_2(0)$ play a role.
Note that the form factors defined this way are dimensionful: 
$G_1$ in GeV$^{-1}$, $G_2$ in GeV$^{-2}$, and $G_3$ in GeV$^{-3}$.
They are analogous to the Pauli-Dirac form factors $F_1$ and $F_2$ for 
the nucleon.

Another definition commonly used in lattice QCD calculations~\cite{Derek93,alex04} is
\begin{equation}
\begin{array}{l}
 < \Delta (k',s')|j^\mu  |N(k,s) >
 \\
  = i\sqrt {\frac{2}{3}} \left ( m_{\Delta} m_N \over E_\Delta E_N \right)^{1/2}
    \bar u_\rho(k',s')\Gamma^{\rho \mu} u(k,s),
\end{array}
\end{equation}
where 
\begin{equation}
\Gamma^{\rho \mu}  = G_{M1} (q^2 )K_{M1}^{\rho \mu}  + G_{E2}
(q^2 )K_{E2}^{\rho \mu}  + G_{C2} (q^2 )K_{C2}^{\rho \mu},
\end{equation}
in terms of the magnetic dipole $G_{M1}$, electric quadrupole $G_{E2}$, and
Coulomb quadrupole $G_{C2}$ form factors. 
They are analogous to the Sachs form factors $G_E$ and $G_M$ for the nucleon.
The invariant functions involve the masses of the particles explicitly,
\begin{equation}
\begin{array}{rcl}
 K_{M1}^{\rho \mu}  &=&  -\frac{3}{{(m_{\Delta}  + m_N )^2  + q^2}}
 \frac{{(m_{\Delta} + m_N )}}{{2m_N}} \varepsilon ^{\rho \mu \alpha \beta} P_\alpha  q_\beta;   \\
 K_{E2}^{\rho \mu} &=& -K_{M1}^{\rho \mu} - 6\Omega^{-1}(q^2)\varepsilon^{\rho \lambda \alpha \beta
}
  P_\alpha  q_\beta  {\varepsilon^\mu \lambda}^{ \gamma \delta}(2 P_\gamma +q_\gamma)
  \\&& \times q_\delta i\gamma _5 \frac{{(m_{\Delta}   + m_N )}}{{2m_N}};   \\
 K_{C2}^{\rho \mu}  &=&  - 6\Omega ^{ - 1} (q^2 )q^\rho( q^2 P^\mu
  - q \cdot Pq^\mu ) i \gamma_5\frac{{(m_{\Delta} + m_N )}}{{2m_N}}, \\
 \end{array}
\end{equation}
where $P=(k'+k)/2$ and 
$\Omega(q^2)=[(m_{\Delta} + m_N )^2 - q^2 ][(m_{\Delta} - m_N )^2 - q^2 ]$.

We stress that the two sets of form factors are equivalent in describing 
the physics of N to $\Delta$ transitions. They are related by
\begin{equation}
\begin{array}{rcl}
 G_{M1} (Q^2 ) &=& \frac{{m_N}}{{3(m_N  + m_{\Delta}  )}}[((3m_{\Delta}   + m_N )(m_{\Delta}   + m_N ) + Q^2 )
   \\&&\frac{{G_1 (Q^2 )}}{{m_{\Delta} }}
 + (m_{\Delta} ^2  - m_N^2 )G_2 (Q^2 ) - 2Q^2 G_3 (Q^2 )] ;\\
 G_{E2} (Q^2 ) &=& \frac{{m_N}}{{3(m_N  + m_{\Delta}  )}}[(m_{\Delta} ^2  - m_N^2  - Q^2 )\frac{{G_1 (Q^2 )}}{{m_{\Delta} }}
 \\&& + (m_{\Delta} ^2  - m_N^2 )G_2 (Q^2 ) - 2Q^2 G_3 (Q^2 )]; \\
 G_{C2} (Q^2 ) &=& \frac{{2m_N}}{{3(m_{\Delta}   + m_N )}}[2m_{\Delta}  G_1 (Q^2 ) + \frac{1}{2}(3m_{\Delta} ^2  + m_N^2  + Q^2 )
 \\&&G_2 (Q^2 ) + (m_{\Delta} ^2  - m_N^2  - Q^2 )G_3 (Q^2 )].
 \end{array}
\label{g1g2}
\end{equation}
Here we have switched to the notation of $Q^2=-q^2$ since most studies of 
form factors are for space-like momentum transfers ($q^2<0$ or $Q^2>0$).
The inverse form, omitting the explicit $Q^2$ dependence in the G's, is given as:
\begin{equation}
\begin{array}{rcl}
  G_{1}&=& -{3 (G_{E2}- G_{M1}) [ m_{\Delta} (m_{\Delta}+ m_N)]}
  \\&& /{[2  m_N  ((m_{\Delta}+m_N)^2+ Q^2 )]};
 \\ G_{2}&=&-3 (m_{\Delta}+ m_N) [ (-2  G_{C2}
   Q^2 \\&& + G_{M1}  (m_{\Delta}^2-2  m_{\Delta}  m_N+ m_N^2+ Q^2 )
   \\&& + G_{E2}  (-3    m_{\Delta}^2+2  m_{\Delta}  m_N+ m_N^2+ Q^2 ) )]
 \\&& /[2  m_N  ((m_{\Delta}^2- m_N^2- Q^2)^2 + 4 m_N^2 Q^2 )];
  \\ G_{3}&=& -3 (m_{\Delta} + m_N) [ (-2  G_{C2}  (m_{\Delta}^2- m_N^2 )
  \\&&  +  G_{M1}  (m_{\Delta}^2-2  m_{\Delta}  m_N + m_N^2+ Q^2 )
   \\&& + G_{E2}  (5  m_{\Delta}^2+2  m_{\Delta}  m_N+ m_N^2+ Q^2 ))]
   \\&& / [{4  m_N  ((m_{\Delta}^2- m_N^2- Q^2)^2 + 4 m_N^2 Q^2
   )}].
\end{array}
\label{gmge}
\end{equation}
From these relations, we see that $G_{M1}$ is on the same order as $G_1$,
$G_{E2}$ is proportional to $G_1/m_\Delta+G_2$ at $Q^2=0$. 
The smallness of $G_{E2}$ comes from cancelation between $G_1$ and $G_2$, 
making it quite difficult to quantify accurately.

The Particle Data Group (PDG)~\cite{PDG08} uses the transition amplitudes 
$f_{M1}$ and $f_{E2}$ 
which are re-scaled versions of the Sachs form factors 
\begin{equation}
\begin{array}{rcl}
 f_{M1} &=& {e\over 2m_N}\left( {|\vec{q}| m_\Delta \over m_N} \right)^{1/2} G_{M1},\\
 f_{E2} &=&-{e\over 2m_N}\left( {|\vec{q}| m_\Delta \over m_N} \right)^{1/2} 
{2|\vec{q}| m_\Delta \over m_\Delta^2-m_N^2} G_{E2},\\
 \end{array}
 \end{equation}
where $e=\sqrt{4\pi\alpha}$.
In the rest frame of the $\Delta$ at $q^2=0$, energy-momentum conservation sets
$2|\vec{q}|m_{\Delta}=m_{\Delta}^2-m_N^2$ so the relations simplify to 
\begin{equation}
\begin{array}{rcl}
 f_{M1} &=& \frac{e}{{2m_N}}
  (\frac{m_{\Delta}^2-m_N ^2}{{2m_N}})^{1/2}{G_{M1}};
  \\
 f_{E2} &=&  - \frac{e}{{2m_N}}(\frac{m_{\Delta}^2-m_N ^2}{{2m_N}})^{1/2}{G_{E2}}.
 \end{array}
 \end{equation}
They are related to the well-known helicity amplitudes by 
\begin{equation}
\begin{array}{l}
 f_{M1} = \frac{{ - 1}}{{2\sqrt 3}}(3A_{3/2}  + \sqrt 3 A_{1/2} );
\\
 f_{E2} = \frac{1}{{2\sqrt 3}}(A_{3/2}  - \sqrt 3 A_{1/2} ).
 \end{array}
 \end{equation}
A commonly quoted quantity is the ratio 
\begin{equation}
R_{EM} = { f_{E2} \over f_{M1} } =  - { G_{E2} \over  G_{M1} }.
\end{equation}

Partial photonuclear decay widths may also be related to the Sachs form factors assuming continuum 
dispersion relations,
\begin{equation}
\begin{array}{l}
 \Gamma_{M1} = {\alpha\over 16} {(m_\Delta^2-m_N^2)^3 \over m_\Delta^3 m_N^2 } G_{M1}^2,
\\
 \Gamma_{E2} = {3\alpha\over 16} {(m_\Delta^2-m_N^2)^3 \over m_\Delta^3 m_N^2 } G_{E2}^2.
 \end{array}
 \end{equation}


\subsection{Experimental information}

Throughout this work, we use the generic term N to $\Delta$ transition to 
refer to the eight transitions from the baryon octet to the decuplet:
\begin{equation}
\begin{array}{lcl}
\gamma p & \rightarrow & \Delta^+,  \\
\gamma n & \rightarrow & \Delta^0,  \\
\gamma \Sigma^+ & \rightarrow & \Sigma^{*+},  \\
\gamma \Sigma^0 & \rightarrow & \Sigma^{*0},  \\
\gamma \Sigma^- & \rightarrow & \Sigma^{*-},  \\
\gamma \Xi^0 & \rightarrow & \Xi^{*0},  \\
\gamma \Xi^- & \rightarrow & \Xi^{*-}, \\
\gamma \Lambda & \rightarrow & \Sigma^{*0}.
\end{array}
\label{octdec}
\end{equation}

Experimentally, pion photo- and or electro-production has been used to access  
the transition amplitudes $G_M$ and $G_E$.
The only transition measured so far is $\gamma p \rightarrow \Delta^+$.
The Laser Electron Gamma Source (LEGS) at Brookhaven has  made 
measurement of proton cross sections and photon asymmetries.
Experiments on pion electroproduction were performed at the Mainz Microtron (MAMI)
in the $\Delta$ resonance region and low momentum transfer. A program
using the CEBAF Large Acceptance Spectrometer (CLAS) at Jefferson
Lab has been inaugurated to improve the systematic and statistical
precision by covering a wide kinematic range of four-momentum
transfer $Q^2$. At the MIT-Bates linear accelerator, an extensive
program has also been developed.

\begin{table}[htb]
\caption{A summary of experimental results for the
ratio $R_{EM}$ in the $\gamma p \rightarrow \Delta^+$ transition.}
\label{expdata} 
\begin{tabular}{lcccc}
\hline\hline Experiment & $Q^2$    &$G_{M1}$  &$G_{E2}$ &  $R_{EM}$\\
          & (GeV$^2$) & ($\mu_N$) & ($\mu_N$) &(\%)    \\
\hline
MAMI~\cite{Sparveris07} & 0.06  & 2.51 & 0.057(14) & -2.28(62)\\
MAMI~\cite{David97} & 0.38 &3.01(1)&0.096(9)  & -3.2(0.25)     \\
LEGS~\cite{Blanpied97} & 0 &&& -2.5 \\
JLab~\cite{Joo02} & 0.4 &  1.16&0.040(10)&  -3.4(4)(4) \\
Bates~\cite{Mertz01} & 0.126 &2.2&0.050(9) &  -2.0(2)(2)  \\
PDG~\cite{PDG08} & 0  &3.02&0.050& -2.5(0.5)\\
\hline\hline
\end{tabular}
\end{table}

Table~\ref{expdata} lists some of the main experimental data and analysis
results at low momentum transfer. It is clear from the results on $R_{EM}$ that
the quadruple component in the N to $\Delta$ transition is quite small,
signaling a slight deformation of the nucleon. The minus sign implies
that its shape is oblate rather than prolate. The bulk of the transition
is dominated by the magnetic dipole component, which results from a
spin-flip in one of the quarks.
The PDG results corresponding to $G_{M1}$ and $G_{E2}$ in other notations are
$G_{1}=2.70$ GeV$^{-1}$, $G_{2}=-1.65$ GeV$^{-2}$;
$A_{1/2}=-0.135$ GeV$^{-1/2}$, $A_{3/2}=-0.250$ GeV$^{-1/2}$;
$f_{M1}=0.284$ GeV$^{-1/2}$, $f_{E2}=-0.0047$ GeV$^{-1/2}$;

The extraction of the small $G_{E2}$ amplitude is very sensitive to
the details of the models and the specific database used in the fit.
Three commonly used are the dynamical model of Sato and Lee (SL
model)~\cite{Sato96}, the unitary isobar model MAID~\cite{Drechsel99}, 
and the partial-wave analysis model SAID~\cite{Igor02}.
The SL model uses an effective Hamiltonian
consisting of bare vertex interactions and energy-independent
meson-exchange $N \to\Delta $ transition operators which are derived by
applying a unitary transformation to a model Lagrangian describing
interactions among $\gamma, \Pi, \rho, \omega, N$ and $\Delta$
fields. With appropriate phenomenological form factors and
coupling constants for $\rho$ and $\Delta$, the model gives a good
description of $\pi N$ scattering phase shifts up to the $\Delta $
excitation energy region.
The MAID and SAID models are partial-wave-based models
for pion photo and electroproduction. It start from the Lagrangian
for the nonresonant terms, including explicit nucleon and light
meson degrees of freedom coupled to the electromagnetic field.
The resonance contributions are included by taking into account unitarity
to provide the correct phases of the pion photoproduction multipoles.
This model provides a good description for individual multipoles as well
as differential cross sections and polarization observables.

\section{QCD Sum Rule Method}
\label{meth}

The starting point is the time-ordered correlation function 
between a nucleon state and a $\Delta$ state in the
QCD vacuum in the presence of a {\em static} background
electromagnetic field $F_{\mu\nu}$:
\begin{equation}
\Pi_\alpha (p)=i\int d^4x\; e^{ip\cdot x} \langle 0\,|\,
T\{\;\eta_\alpha^{\Delta}(x)\, \bar{\eta}^{N}(0)\;\}\,|\,0\rangle_F,
\label{cf2pt}
\end{equation}
where $\eta_\alpha^{\Delta}$ ($\alpha$ Lorentz index) 
and $\eta^{N}$ are interpolating field operators 
carrying the quantum numbers of the baryons under consideration.
The task is to evaluate this correlation on two
different levels. On the quark level, the correlation function describes a hadron as
quarks and gluons interacting in the QCD vacuum. On the
phenomenological level, it is saturated by a tower of hadronic
intermediate states with the same quantum numbers. By matching the two, a
link can be established between a description in terms of
hadronic degrees of freedom and one based on the underlying quark
and gluon degrees of freedom as governed by QCD. 
The subscript $F$ means that
the correlation function is to be evaluated with an
electromagnetic interaction term,
\begin{equation}
{\cal L}_I = - A_\mu j^\mu,
\end{equation}
 added to the QCD Lagrangian.
Here $A_\mu$ is the external electromagnetic potential and
$j^\mu=e_q \bar{q} \gamma^\mu q$ is the quark electromagnetic current.

Since the external field can be made arbitrarily small, one can
expand the correlation function
\begin{equation}
\Pi_\alpha (p)=\Pi_\alpha^{(0)}(p) +\Pi_\alpha^{(1)}(p)+\cdots,
\end{equation}
where $\Pi_\alpha^{(0)}(p)$ is the correlation function in the absence of
the field, and gives rise to the mass sum rules of the hadron.
The transition amplitudes will be extracted from the linear response function $\Pi_\alpha^{(1)}(p)$.
The action of the external electromagnetic field is two-fold: on the one hand it
couples directly to the quarks in the hadron interpolating fields;
on the other it polarizes the QCD vacuum. The latter can be described
by introducing new parameters called QCD vacuum susceptibilities.

\subsection{Interpolating fields}\label{inte}

We need interpolating fields for the baryons in N to $\Delta$ transitions 
in Eq.~(\ref{octdec}). They are built from quark field operators 
with appropriate quantum numbers.
For the nucleon (a spin-1/2 object) it is not unique.
We consider a linear combination of the two standard local
interpolating fields for the baryon octet:
\begin{equation}
\begin{array}{lcl}
\eta^{p}(uud)&=&-2\epsilon^{abc} [(u^{aT}C\gamma_5 d^b)u^c+\beta
(u^{aT}C d^b)\gamma_5 u^c],
\\
\eta^{n}(ddu)&=&-2\epsilon^{abc} [(d^{aT}C\gamma_5 u^b)d^c+\beta
(d^{aT}C u^b)\gamma_5 d^c],
\\
\eta^{\Lambda}(uds)&=&{-2\sqrt{\frac{1}{6}}}\epsilon^{abc} [
2(u^{aT}C\gamma_5 d^b)s^c +(u^{aT}C\gamma_5 s^b)d^c
\\&& -(d^{aT}C\gamma_5 s^b)u^c \hspace{2mm}
 +\beta(2(u^{aT}C d^b)\gamma_5s^c \\&& +(u^{aT}C s^b)\gamma_5d^c  -(d^{aT}C
s^b)\gamma_5u^c )],
\\
\eta^{\Sigma^+}(uus)&=&-2\epsilon^{abc} [(u^{aT}C\gamma_5
s^b)u^c+\beta  (u^{aT}C s^b)\gamma_5 u^c],
\\
\eta^{\Sigma^0}(uds)&=&-\sqrt{2}\epsilon^{abc} [
 (u^{aT}C\gamma_5 s^b)d^c+(d^{aT}C\gamma_5 s^b)u^c
\\&&
+\beta((u^{aT}C s^b)\gamma_5d^c +(d^{aT}C s^b)\gamma_5u^c) ],
\\
\eta^{\Sigma^-}(dds)&=&-2\epsilon^{abc} [(d^{aT}C\gamma_5
s^b)d^c+\beta (d^{aT}C s^b)\gamma_5 d^c],
\\
\eta^{\Xi^0}(ssu)&=&-2\epsilon^{abc} [(s^{aT}C\gamma_5
u^b)s^c+\beta (s^{aT}C u^b)\gamma_5 s^c],
\\
\eta^{\Xi^-}(ssd)&=&-2\epsilon^{abc} [(s^{aT}C\gamma_5
d^b)s^c+\beta (s^{aT}C d^b)\gamma_5 s^c].
\end{array}
\label{octinte}
\end{equation}
Here $u$, $d$, and $s$ are up, down, and strange quark field operators, $C$
is the charge conjugation operator, the superscript $T$ means
transpose, and $\epsilon_{abc}$ ensures that the constructed baryons 
are color-singlet. The real
parameter $\beta$ allows for the mixture of the two independent
currents. The choice advocated by Ioffe~\cite{Ioffe81} and often
used in QCD sum rules studies corresponds to $\beta=-1$. In this
work, $\beta$ is allowed to vary in order to achieve maximal overlap with the
state in question for a particular sum rule. The normalization
factors are chosen so that in the limit of SU(3)-flavor symmetry
all N to N correlation functions simplify to that of the proton.

For the spin-3/2 baryon decuplet, we use
\begin{equation}
 \begin{array}{lcl}
   \eta_\alpha^{\Delta^+}(uud)&=&{1\over\sqrt{3}} \epsilon^{abc} [2(u^{aT}C\gamma_\alpha d^b) u^c +(u^{aT}C\gamma_\alpha u^b) d^c ],
   \\ \eta_\alpha^{\Delta^0}(ddu)&=&{1\over\sqrt{3}} \epsilon^{abc} [2(d^{aT}C\gamma_\alpha u^b) d^c +(d^{aT}C\gamma_\alpha d^b) u^c ],
 \\ \eta_\alpha^{\Sigma^{*+}}(uus)&=&{1\over\sqrt{3}} \epsilon^{abc} [2(u^{aT}C\gamma_\alpha s^b) u^c +(u^{aT}C\gamma_\alpha u^b) s^c ],
    \\ \eta_\alpha^{\Sigma^{*0}}(uds)&=&{2\over\sqrt{3}} \epsilon^{abc} [ (u^{aT}C\gamma_\alpha d^b) s^c+(d^{aT}C\gamma_\alpha s^b) u^c
    \\&& +(s^{aT}C\gamma_\alpha u^b) d^c ],
   \\ \eta_\alpha^{\Sigma^{*-}}(dds)&=&{1\over\sqrt{3}} \epsilon^{abc} [2(d^{aT}C\gamma_\alpha s^b) d^c +(d^{aT}C\gamma_\alpha d^b) s^c ],
 \\ \eta_\alpha^{\Xi^{*0}}(ssu)&=&{1\over\sqrt{3}} \epsilon^{abc} [2(s^{aT}C\gamma_\alpha u^b) s^c +(s^{aT}C\gamma_\alpha s^b) u^c
 ],
 \\ \eta_\alpha^{\Xi^{*-}}(ssd)&=&{1\over\sqrt{3}} \epsilon^{abc} [2(s^{aT}C\gamma_\alpha d^b) s^c +(s^{aT}C\gamma_\alpha s^b) d^c
 ].
 \end{array}
 \label{decinte}
\end{equation}

\begin{widetext}

\subsection{Phenomenological representation}
\label{rhs}

We begin with the structure of the correlation function in the
presence of the electromagnetic vertex to first order:
\begin {equation}
\Pi _\alpha  (p) = i\int {d^4 x} e^{ipx}  < 0|\eta _\alpha ^\Delta
(x)[ - i\int {d^4 y} A_\mu  (y)j^\mu  (y)]\bar \eta ^{\rm N} (0)|0 >.
\end {equation}
After inserting two complete sets of physical intermediate states,  it becomes
\begin {equation}
\begin{array}{l}
 \Pi _{\alpha}  (p) = \int {d^4 x} d^4 y\frac{{d^4 k'}}{{(2\pi )^4 }}\frac{{d^4 k}}{{(2\pi )^4 }}\sum\limits_{\Delta N} \sum\limits_{s's}
 {\frac{{ - i}}{{k'^2  - m_{\Delta} ^2  - i\varepsilon }}}  \frac{{ - i}}{{k^2  - m_N^2  - i\varepsilon }}
e^{ipx} \\ \times A_\mu (y) < 0|\eta ^\Delta_\alpha (x)|\Delta k's' >  <
 \Delta k's'|j^\mu  (y)| Nks> <Nks|\bar \eta (0)|0 >.
\end{array}
\end {equation}
Using the translation invariance on $\eta^\Delta_\alpha(x)$ and $ j^\mu(y)$,
and the overlap strengths ($\lambda_\Delta$ and $\lambda_N$) 
of the interpolating fields with the states defined by,
\begin {equation}
\begin{array}{l}
  < 0|\eta _\alpha  (0)|\Delta k',s' >  = \lambda _\Delta  u _\alpha  (k',s' ), \\
  < Nks |\eta (0)|0 >  = \lambda _N^* \bar u (k,s ),
 \end{array}
\end {equation}
we can write out explicitly the contribution of the lowest lying
state and designate by ESC the excited state contributions:
\begin {equation}
\begin{array}{l}
 \Pi _\alpha  (p) =  - \lambda _\Delta  \lambda _N^* \int {d^4 x} d^4 y\frac{{d^4 k'}}{{(2\pi )^4 }}\frac{{d^4 k}}{{(2\pi )^4 }}[k'^2  - m_{\Delta} ^2  - i\varepsilon ]^{ - 1} [k^2  - m_N^2  - i\varepsilon ]^{ - 1}  \\
 A_\mu  (y)e^{i(p - k')x} e^{iqy} \sum\limits_{s'}  {u_\alpha  (k',s' )\bar u _\rho  (k',s' )\Gamma^{\rho \mu } } \sum\limits_{s } {u (k,s )\bar u (k,s )}  + ESC.
 \end{array}
\end {equation}
The spinor sum for $N$ in the above expression is 
\begin {equation}
\sum\limits_{s} {u(k,s)\bar u(k,s  )}  = \hat k + m_N,
\end {equation}
and the spinor sum for $\Delta$ is 
\begin {equation}
 \sum\limits_{s'}  {\bar u _\alpha  (k',s' )u _\rho  (k',s' ) =  - (\hat k' + m_{\Delta}  )\{ g_{\alpha \rho }  - \frac{1}{3}\gamma _\alpha  \gamma _\rho   - \frac{{2k'_\alpha  k'_\rho  }}{{3m_{\Delta} ^2 }} + \frac{{k'_\alpha  \gamma _\rho   - k'_\rho  \gamma _\alpha  }}{{3m_{\Delta}  }}}
 \}.
\end {equation}
We work in the fixed-point gauge in which $A_\mu (y) = - \frac{1}{2}F_{\mu \nu } y^\nu$.
After changing variables from $k$ to $q=k'-k$, which leads to $d^4 k =- d^4 q$, 
we have 
\begin {equation}
\begin{array}{l}
 \Pi_\alpha (p) =  - \lambda _N^* \lambda _\Delta  F_{\mu \nu } \int {d^4 x} d^4 y\frac{{d^4 k'}}{{(2\pi )^4 }}\frac{{d^4 q}}{{(2\pi )^4 }}[k'^2  - m_{\Delta} ^2][(q - k')^2  - m_N^2 ]^{-1}
 \\\times e^{i(p - k')x} (- {i\over 2}\frac{\partial }{{\partial q_\nu  }}e^{iqy} )(\hat k' + m_{\Delta}  )\{ g_{\alpha \rho }  - \frac{1}{3}\gamma _\alpha  \gamma _\rho   - \frac{{2k'_\alpha  k'_\rho  }}{{3m_{\Delta} ^2 }} + \frac{{k'_\alpha  \gamma _\rho   - k'_\rho  \gamma_\alpha }}{{3m_{\Delta}  }}\}\times \Gamma^{\rho \mu } \times (\hat {k'} - \hat q + m_N ).
 \end{array}
\end {equation}
Integrating by parts and using properties of $\delta$ functions, the expression collapses to 
\begin {equation}
\begin {array}{rcl}
\Pi _\alpha  (p) &=& \frac{{  i\lambda _N^* \lambda _\Delta
}}{2}F_{\mu \nu } [p^2  - m_{\Delta} ^2]^{ - 1} [p^2 - m_N^2 ]^{ -
1} (\hat p + m_{\Delta} )
\\&&\times\{ g_{\alpha
\rho }  - \frac{1}{3}\gamma _\alpha  \gamma _\rho   -
\frac{{2p_\alpha  p_\rho  }}{{3m_{\Delta} ^2 }} - \frac{{p_\alpha
\gamma _\rho   - p_\rho  \gamma _\alpha  }}{{3m_\Delta }}\}
\\&&\times[G_1
(g_{\rho \nu } \gamma _\mu   - \gamma _\nu  g_{\rho \mu }
)\gamma _5  + G_2(g_{\rho \nu } p_\mu   - p_\nu  g_{\rho
\mu } )\gamma _5 ](\hat p + m_N ),
\end{array}
\end {equation}
where we have chosen to work with the transition amplitudes
$G_1$ and $G_2$ as defined in Eq.~(\ref{js1}) and Eq.~(\ref{js2}), 
rather than $G_{M1}$ and $G_{E2}$. 
They are readily converted back and forth by Eq.~(\ref{g1g2}) and Eq.~(\ref{gmge}).  

A straightforward evaluation of this expression leads to numerous
tensor structures, but not all of them are independent of each
other. The dependencies can be removed by ordering the gamma
matrices in a specific order. 
After a lengthy
calculation we identified 12 independent tensor structures. 
They can be organized (aside from an overall factor of $i$ that also appears on the QCD side) as:
\begin {equation}
\begin{array}{l}
 \Pi _\alpha  (p) = \frac{{ \lambda _N^* \lambda _\Delta  }}{2}  
[ p^2  - m_{\Delta} ^2]^{ - 1} [p^2 - m_N^2 ]^{ - 1}\\
 {\rm{[WE_1(}}\frac{2}{3}G_1 m_{\Delta}  {+}\frac{2}{3}G_1 m_N
 {\rm{)}}\\
{\rm{ + WE_2(}}\frac{2}{3}p^2 G_1 m_{\Delta}  ^{-1} {\rm{ - 4/3G}}_1 m_{\Delta}  {\rm{ - }}\frac{2}{3}G_1 m_N {\rm{)}} \\
 {\rm{ + WE_3(}}\frac{2}{3}G_1 m_{\Delta}  {+}\frac{2}{3}G_1 m_N
 {\rm{)}}\\
 {\rm{ + WE_4(- }}\frac{2}{3}p^2 G_1 m_{\Delta}  ^{-1} {+}\frac{2}{3}p^2 G_2 {+}\frac{8}{3}G_1 m_{\Delta}  {+}\frac{2}{3}G_1 m_N {\rm{ - }}\frac{2}{3}G_2 m_{\Delta}  m_N {\rm{)}} \\
 {\rm{ + WE_5(- }}\frac{4}{3}G_1 m_{\Delta}  ^{{\rm{ - 2}}} m_N {\rm{ - }}\frac{2}{3}G_2 m_{\Delta}  ^{-1} m_N {+}\frac{2}{3}G_2
 {\rm{)}}\\
 {\rm{ + WE_6(2p}}^2 G_2 {\rm{ - 2G}}_2 m_{\Delta}  m_N {\rm{)}} \\
 {\rm{ + WO_1(- }}\frac{2}{3}p^2 G_1 {\rm{ - }}\frac{2}{3}G_1 m_{\Delta}  m_N {\rm{)}} \\
 {\rm{ + WO_2(}}\frac{2}{3}G_1 m_{\Delta}  ^{-1} m_N {+}\frac{2}{3}G_1
 {\rm{)}}\\
 {\rm{ + WO_3(- }}\frac{2}{3}p^2 G_1 {\rm{ - }}\frac{2}{3}G_1 m_{\Delta}  m_N {\rm{)}} \\
 {\rm{ + WO_4(}}\frac{4}{3}p^2 G_1 m_{\Delta}  ^{{\rm{ - 2}}} {+}\frac{2}{3}p^2 G_2 m_{\Delta}  ^{-1} {+}\frac{4}{3}G_1 m_{\Delta}  ^{-1} m_N {\rm{ - }}\frac{4}{3}G_1 {\rm{ - }}\frac{4}{3}G_2 m_{\Delta}  {+}\frac{2}{3}G_2 m_N
 {\rm{)}}\\
 {\rm{ + WO_5(- }}\frac{2}{3}G_1 m_{\Delta}  ^{-1} m_N {\rm{ - 2G}}_1 {\rm{ - }}\frac{2}{3}G_2 m_{\Delta}  {+}\frac{2}{3}G_2 m_N {\rm{)}} \\
 {\rm{ + WO_6(}}\frac{4}{3}G_1 m_{\Delta}  ^{-1} m_N {\rm{ - }}\frac{2}{3}G_2 m_{\Delta}  {+}\frac{2}{3}G_2 m_N
 {\rm{)]}}.\
 \end{array}
\label{wp}
\end {equation}

The tensor structures associated with $WE_i$  and $WO_i$ are
listed as follows:
\begin{equation}
\begin{array}{*{20}c}
   \begin{array}{l}
 WE_1 = F_{\mu \nu}\gamma _5 \hat p\gamma _\mu  \gamma _\nu  \gamma _\alpha  \\
 WE_2 =F_{\mu \nu} \gamma _5 \gamma _\mu  \gamma _\nu  p_\alpha  \\
 WE_3 = F_{\mu \nu}\gamma _5 \hat p(\gamma _\nu  g_{\alpha\mu } - \gamma _\mu  g_{\alpha\nu })  \\
 WE_4 =F_{\mu \nu}\gamma _5 (\gamma _\mu  \gamma _\alpha p_\nu - \gamma _\nu  \gamma _\alpha p_\mu )  \\
 WE_5 = F_{\mu \nu} \gamma _5 \hat p (\gamma _\mu  p_\nu  p_\alpha -\gamma _\nu  p_\mu  p_\alpha ) \\
 WE_6 =F_{\mu \nu} \gamma _5 (g_{\alpha\mu } p_\nu - g_{\alpha\nu } p_\mu)   \\
 \end{array} & \begin{array}{l}
 WO_1 = F_{\mu \nu}\gamma _5 \gamma _\mu  \gamma _\nu  \gamma _\alpha  \\
 WO_2 = F_{\mu \nu}\gamma _5 \hat p\gamma _\mu  \gamma _\nu  p_\alpha  \\
 WO_3 =F_{\mu \nu} \gamma _5 (\gamma _\nu  g_{\alpha\mu } - \gamma _\mu  g_{\alpha\nu })  \\
 WO_4 = F_{\mu \nu} \gamma _5 (\gamma _\mu  p_\nu - \gamma _\nu  p_\mu)  p_\alpha  \\
 WO_5 = F_{\mu \nu} \gamma _5 \hat p (\gamma _\mu  \gamma _\alpha p_\nu - \gamma _\nu  \gamma _\alpha p_\mu  ) \\
 WO_6 =F_{\mu \nu} \gamma _5 \hat p (g_{\alpha\mu} p_\nu - g_{\alpha\nu} p_\mu )  \\
 \end{array}  \\
\end{array}
\label{wewo}
\end {equation}
The naming convention of $WE_i$ and $WO_i$ is motivated by even-odd considerations. 
It turns out that the QCD sum rules associated with $WE_i$ will have even-dimension vacuum condensates, 
while the ones associated with $WO_i$ will have odd-dimension ones. Interestingly, 
the $WE_i$ structures themselves contain odd number of gamma matrices, while $WO_i$ even number.

The next step is to perform the standard Borel transform defined  by
\begin{equation}
\hat{B}[f(p^2)] = \mathop{\lim}_{\scriptstyle {-p^2, n \rightarrow
\infty}\hfill\atop \scriptstyle {-p^2/n=M^2}\hfill} \frac{1}{n!}
(-p^2)^{n+1}(\frac{d}{dp^2})^n f(p^2), \label{borel_trans}
\end{equation}
which turns a function of $p^2$ to a function of $M^2$ ($M$ is the Borel mass). 
The transform is needed to suppress excited-state contributions relative to the ground-state.
Unlike the N to N transition in the calculation of 
magnetic moments~\cite{Wang08}, the double-pole structure in the N to $\Delta$  transition in Eq.~(\ref{wp}) has unequal masses $m_\Delta$ and $m_N$. 
The existence of an equal-mass double pole in the ground state 
was crucial in isolating the magnetic moments from the excited states because it gives a 
different functional dependence on the Borel mass ($e^{-m^2/M^2}/M^2$) 
than the single poles ($e^{-m^2/M^2}$). 
Here for N to $\Delta$ transitions, 
we shall approximate the unequal-mass double pole by an equal-mass double pole,
\begin {equation}
\frac{1}{p^2 - m_{N}}\frac{1}{p^2 - m_{\Delta}}\approx(\frac{1}{p^2 - \bar m ^2})^2,
\end {equation}
where $\bar m^2  = ({{m_N ^2  + m_{\Delta}  ^2 }})/{2}$ can be considered as 
an averaged square mass.
For the physical masses involved in the N to $\Delta$ transitions, this approximation 
is good to within $3\%$ in the Borel region of interest we are going to explore.
Because of this approximation, the determination of the N to $\Delta$ transition 
amplitudes will not be as accurate as the magnetic moments in the QCD sum rule approach.
With the introduction of the pure double pole, the Borel transform of the ground state 
in Eq.~(\ref{wp}) is given by 
\begin{equation}
\begin{array}{l}
 \Pi_\alpha  (p) = \frac{{ \lambda _N^* \lambda _\Delta  }}{2M^2} e^{-\bar
m^2/M^2}\\
\rm{[} WE_1 \frac{2}{3}G_1(m_{\Delta} + m_N) \\
 +WE_2 \frac{2}{3}G_1 \frac{1}{m_{\Delta}}(-M^2+\bar m ^2 - 2 m_{\Delta}^2 -
m_N m_{\Delta}) \\
 +WE_3\frac{2}{3}G_1(m_{\Delta} + m_N) \\
 -WE_4 (\frac{2}{3}G_1 \frac{1}{m_{\Delta}}( -M^2+\bar m ^2 -4 m_{\Delta}^2
-m_N m_{\Delta}) + \frac{2}{3}G_2(-M^2+\bar m ^2 -  m_{\Delta} m_N) )\\
 - WE_5(\frac{4}{3}G_1 \frac{1}{m_{\Delta} ^2}m_N + \frac{2}{3}G_2
\frac{1}{m_{\Delta}}(- m_N +  m_{\Delta}) )\\
 +WE_6 2G_2(-M^2+\bar m ^2 -  m_{\Delta} m_N) \\
 -WO_1 \frac{2}{3}G_1(-M^2+\bar m ^2 +  m_{\Delta} m_N) \\
 +WO_2\frac{2}{3}G_1 \frac{1}{m_{\Delta}}(m_N +  m_{\Delta}) \\
 - WO_3\frac{2}{3}G_1(-M^2+\bar m ^2 +  m_{\Delta} m_N) \\
 + WO_4(\frac{4}{3}G_1 \frac{1}{m_{\Delta} ^2}(-M^2+\bar m ^2 + m_N
m_{\Delta} -  m_{\Delta}^2) + \frac{2}{3}G_2 \frac{1}{m_{\Delta}}(-M^2+\bar
m ^2 - 2 m_{\Delta}^2 + m_N m_{\Delta})) \\
 -WO_5(\frac{2}{3}G_1 \frac{1}{m_{\Delta}}(m_N +  m_{\Delta}) -
\frac{2}{3}G_2(m_{\Delta} - m_N) )\\
 +WO_6 (\frac{4}{3}G_1 \frac{1}{m_{\Delta}}m_N - \frac{2}{3}G_2(m_{\Delta} -
m_N) )\rm{]},
 \end{array}
\end{equation}
where $M$ is the Borel mass parameter, not to be confused with the particle masses $m_N$, 
$m_\Delta$, or $\bar{m}$.

The excited states must be treated with care. 
The pole structure can be written in the generic form 
\begin {equation}
\begin{array}{l}
  \frac{C_1 }{(p^2 -m_N^2 )(p^2  - m_{\Delta}^2 )}
  + \sum\limits_{N^* } {\frac{C_2 }{{(p^2 - m_{N^*}^2 )(p^2  - m_{\Delta
  }^2)}}}
   + \sum\limits_{\Delta^* } {\frac{C_3 }{{(p^2  - m_{N}^2} )(p^2  -
   m_{\Delta^*}^2)}}
  + \sum\limits_{N^*, \Delta^* } {\frac{C_4 }{{(p^2  - m_{N^*}^2} )(p^2  -
  m_{\Delta^*}^2)}},
\end{array}
\end {equation}
where $C_i$ are constants.
 The first term is
 the ground state pole which contains the desired amplitudes $G_1$ and $G_2$. 
The second and third terms represent the non-diagonal
 transitions between the ground state and the excited states caused
 by the external field. The forth term is pure excited-state to excited-state transitions.
These different contributions can be represented by the diagrams
in Fig.~\ref{transition}.
 %
\begin{figure}
\centerline{\psfig{file=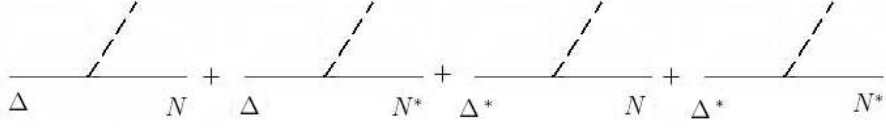,width=12.0cm}}
\vspace*{-15.0cm} \caption{\small{The four contributions to the
spectral function in the presence of an external field: transition
between ground state N to ground state $\Delta$, transitions
between ground state N and excited states $\Delta^*$, transition
between excited state $N^*$ and ground state $\Delta$ and pure
excited states $N^* \to \Delta^*$.}} \label{transition}
\end{figure}
%
Upon Borel transform, it takes the form
\begin {equation}
\begin{array}{l}
 \frac{C_1}{M^2}e^{-{\bar m}^2/M^2}
  +\sum \limits_{N^*,\Delta^* } \frac{C_4}{M^2}e^{-\bar m^{*2}/{M^2}}\;\;\;\;\;\;\;\;\;\;\;\;
  \\+e^{-\bar m^2/M^2} \left[ e^{(\bar m^2-m_{\Delta}^2)/M^2}\sum\limits_{N^* }  
{C_2 (1-e^{-(m_{\Delta}^2-m_{N^*}^2)/M^2}) \over m_{\Delta}^2-m_{N^*}^2 }
  +e^{(\bar m^2-m_{N}^2)/M^2}\sum\limits_{\Delta^* }  
{C_3 (1-e^{-(m_{\Delta^*}^2-m_{N}^2)/M^2}) \over m_{\Delta^*}^2-m_{N}^2 }. \right]
\end{array}
\end {equation}

\end{widetext}
 The important point is that the non-diagonal transitions 
($N^*$ to $\Delta$ and $N$ to $\Delta^*$) give rise to a
contribution that is not  exponentially suppressed relative to the
ground state. This is a general feature of the external-field technique. 
The strength of
 such transitions at each structure is {\it a priori} unknown and is a
potential source of contamination in the determination of the amplitudes.
 The standard treatment of the off-digonal transitions is 
to approximate it (the quantity in the square brackets)
 by a parameter A, which is to be extracted from the sum rule along
 with the ground state property of interest. Inclusion of such
 contributions is necessary for the correct extraction of the
 amplitudes. The pure excited state transitions ($N^*$ to $\Delta^*$) are
 exponentially suppressed relative to the ground state and can be
 modeled in the usual way by introducing a continuum model and
 threshold parameter $w$.

In summary, we identified 12 different tensor structures 
from which 12 sum rules can be constructed, once 
the correlation function is evaluated at the quark level.

\subsection{Calculation of the QCD side}
\label{lhs}
We start by writing down the master formulas used
in the calculation of the QCD side, which are functions of the
quark propagators, obtained by contracting out pairs of
time-ordered quark-field operators in the two-point correlation
function in Eq.~(\ref{cf2pt}) using the interpolating fields 
in Eq.~(\ref{octinte}) and Eq.~(\ref{decinte}). 
We note that the same master formulas enter lattice QCD
calculations where the fully-interacting quark propagators are
generated numerically, instead of the analytical ones used here. 
The master formula for the transition $ \gamma p \to
\Delta^{+} $ (with $uud$ quark content) is:
\begin {equation}
\begin{array}{l}
  < 0|T\{ \eta_\alpha^{\Delta^+} (x)\bar \eta ^p (0)\} |0 >
  =- \frac{4}{{\sqrt 3}}  \varepsilon ^{abc} \varepsilon ^{a'b'c'} \{  \\
 S_u ^{aa'} \gamma _5 CS_d^{bb'^T} C\gamma _\alpha  S_u ^{cc'}
  + S_u ^{aa'} Tr(\gamma _5 CS_u ^{T bb'} C\gamma _\alpha  S_d^{cc'} )
 \\ - S_d^{aa'} \gamma _5 CS_u ^{T bb'} C\gamma _\alpha  S_u ^{cc'}
  + \beta [  S_u ^{aa'} CS_d^{Tbb'}  C\gamma _\alpha  S_u ^{cc'} \gamma _5
  \\+ S_u ^{aa'} \gamma _5 Tr(CS_u ^{T bb'} C\gamma _\alpha S_d^{cc'} )
  - S_d^{aa'} CS_u ^{T bb'} C\gamma _\alpha  S_u ^{cc'}
  ]\}.\label{masterp}
 \end{array}
\end {equation}
It is not necessary to list all the master formulas for the other
transitions separately. Only three are distinct;
others can be obtained by making appropriate substitutions
in the above expression as specified below:
    \begin{itemize}
    \item for $n \gamma \to \Delta^{0}$ (with $ddu$ quark content),
    interchange d quark and u quark in $p\gamma \to \Delta^{+}$,
    \item for $\Sigma^+ \gamma \to \Sigma^{*+}$ (with $uus$ quark content),
    replace d quark by s quark in $p \gamma \to \Delta^{+}$,
    \item for $\Sigma^- \gamma \to \Sigma^{*-}$ (with $dds$ quark content),
    replace u quark by d quark in $\Sigma^+ \gamma \to \Sigma^{*+}  $,
    \item for $\Xi^0 \gamma \to \Xi^{*0}$ (with $ssu$ quark content),
    interchange u quark and s quark in $\Sigma^+ \gamma \to \Sigma^{*+}$,
    \item for $\Xi^- \gamma \to
\Xi^{*-}$ (with $ssd$ quark content), replace u quark by d quark
in $\Xi^0 \gamma \to \Xi^{*0}$.
    \end{itemize}
Here the interchange of two quarks is achieved by simply
switching the corresponding propagators.

The master formula for $\Sigma^0 \gamma \to \Sigma^{*0}$ (with
$uds$ quark content) is:
\begin {equation}
\begin{array}{l}
  < 0 |T\{ \eta_\alpha ^{\Sigma ^{*0}} (x)\bar \eta ^{\Sigma ^0} (0)\} |0 >
   =  \frac{2}{{\sqrt 3}}  \varepsilon ^{abc} \varepsilon ^{a'b'c'} \{  \\
 S_s^{aa'} \gamma _5 CS_u^{T bb'} C\gamma _\alpha  S_d^{cc'}
 - S_u^{aa} \gamma _5 CS_s^{T bb'} C\gamma _\alpha  S_d^{cc'}
  \\- S_d^{aa'} Tr(\gamma _5 CS_s^{Tbb'} C\gamma _\alpha  S_u^{cc'} )
  + S_s^{aa'} \gamma _5 CS_d^{T bb'} C\gamma _\alpha  S_u^{cc'}
 \\ - S_d^{aa} \gamma _5 CS_s^{T bb'} C\gamma _\alpha  S_u^{cc'}
  - S_u^{aa'} Tr(\gamma _5 CS_s^{Tbb'} C\gamma _\alpha  S_d^{cc'} )
 \\ + \beta [S_s^{aa'} CS_u^{T bb'} C\gamma _\alpha  S_d^{cc'} \gamma _5
  - S_u^{aa'} CS_s^{T bb'} C\gamma _\alpha S_d^{cc'} \gamma _5
 \\ - S_d^{aa'} \gamma _5 Tr(CS_s^{T bb'} C\gamma _\alpha S_u^{cc'} )
  + S_s^{aa'} CS_d^{T bb'} C\gamma _\alpha  S_u^{cc'} \gamma _5
  \\- S_d^{aa'} CS_s^{T bb'} C\gamma _\alpha S_u^{cc'} \gamma _5
  - S_u^{aa'} \gamma _5 Tr(CS_s^{T bb'} C\gamma _\alpha S_d^{cc'} )].
 \end{array}\label{mastersig0}
\end {equation}
The master formula for $\Lambda\gamma \to \Sigma^{*0}$ (with $uds$
quark content) is:
\begin {equation}
\begin{array}{l}
  <0 |T\{ \eta_\alpha ^{\Sigma ^{*0}} (x)\bar \eta ^{\Lambda} \} |0 >
   =  \frac{2}{3}   \varepsilon ^{abc} \varepsilon ^{a'b'c'} \{  \\
2 S_d^{aa'} \gamma _5 CS_u^{T bb'} C\gamma _\alpha  S_s^{cc'} -
2S_u^{aa} \gamma _5 CS_d^{T bb'} C\gamma _\alpha  S_s^{cc'}
\\- 2S_s^{aa'} Tr(\gamma _5 CS_u^{Tbb'} C\gamma _\alpha  S_d^{cc'} )
  + S_s^{aa'} \gamma _5 CS_u^{T bb'} C\gamma _\alpha  S_d^{cc'}
 \\  - S_u^{aa} \gamma _5 CS_s^{T bb'} C\gamma _\alpha  S_d^{cc'}
   - S_d^{aa'} Tr(\gamma _5 CS_s^{Tbb'} C\gamma _\alpha  S_u^{cc'} )
  \\- S_s^{aa'} \gamma _5 CS_d^{T bb'} C\gamma _\alpha  S_u^{cc'}
   + S_d^{aa} \gamma _5 CS_s^{T bb'} C\gamma _\alpha  S_u^{cc'}
   \\+ S_u^{aa'} Tr(\gamma _5 CS_s^{Tbb'} C\gamma _\alpha  S_d^{cc'} )
  + \beta [ -2 S_d^{aa'} CS_u^{T bb'} C\gamma _\alpha  S_s^{cc'} \gamma _5
   \\+2 S_u^{aa} CS_d^{T bb'} C\gamma _\alpha  S_s^{cc'} \gamma _5
   - 2S_s^{aa'} Tr(CS_u^{Tbb'} C\gamma _\alpha  S_d^{cc'} )\gamma _5
  \\+ S_s^{aa'} CS_u^{T bb'} C\gamma _\alpha  S_d^{cc'}\gamma _5
  - S_u^{aa} CS_s^{T bb'} C\gamma _\alpha  S_d^{cc'} \gamma _5
  \\- S_d^{aa'} Tr(CS_s^{Tbb'} C\gamma _\alpha  S_u^{cc'} )\gamma _5
  - S_s^{aa'} CS_d^{T bb'} C\gamma _\alpha  S_u^{cc'}\gamma _5
  \\+ S_d^{aa} CS_s^{T bb'}  C\gamma _\alpha  S_u^{cc'} \gamma _5
  + S_u^{aa'} Tr(CS_s^{Tbb'} C\gamma _\alpha  S_d^{cc'} )\gamma _5]\}.
 \end{array}\label{masterlam}
\end {equation}
In the above equations,
\begin{equation}
S^{ab}_q (x,0;F) \equiv \langle 0\,|\, T\{\;q^a(x)\,
\bar{q}^b(0)\;\}\,|\,0\rangle_F,
 \hspace{3mm} q=u, d, s,
\end{equation}
is the fully interacting quark propagator in the presence of the
electromagnetic field. It is obtained by the operator product expansion (OPE) 
 and is made of dozens of analytic terms. 
It has been discussed to various degree in the literature~\cite{Ioffe84,Pasupathy86,Wilson87,Wang08} so we will not list the terms here.

In addition to the standard vacuum condensates, the vacuum
susceptibilities induced by the external field are defined by
\begin{equation}
\begin{array}{rcl}
\langle\bar{q} \sigma_{\mu\nu} q\rangle_F &\equiv&
e_q \chi \langle\bar{q}q\rangle F_{\mu\nu}, \\
\langle\bar{q} g_c G_{\mu\nu} q\rangle_F &\equiv&
e_q \kappa \langle\bar{q}q\rangle F_{\mu\nu}, \\
\langle\bar{q} g_c \epsilon_{\mu\nu\rho\lambda} G^{\rho\lambda}
\gamma_5 q\rangle_F &\equiv &i e_q \xi \langle\bar{q}q\rangle
F_{\mu\nu}.
\end{array}
\end{equation}
Note that $\chi$ has the dimension of GeV$^{-2}$, while $\kappa$
and $\xi$ are dimensionless.
%
\begin{figure}[htb]
\centerline{\epsfig{file=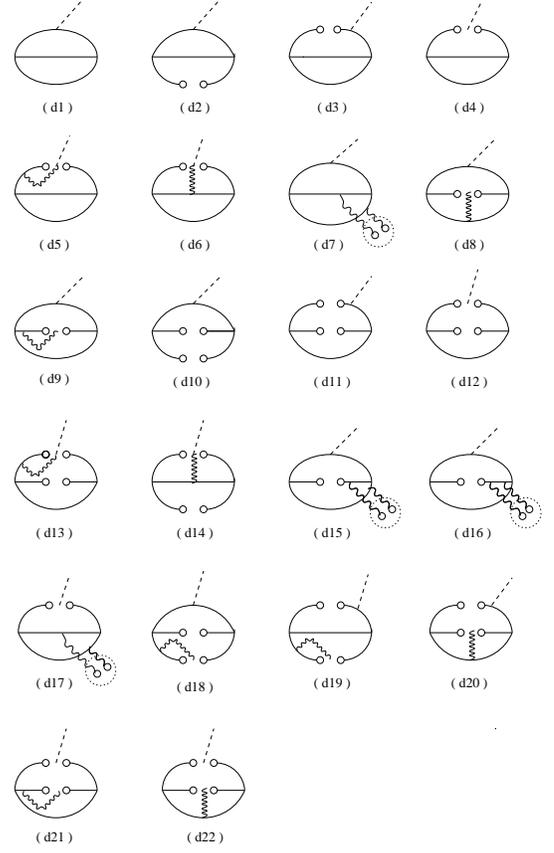,width=7.0cm}} \vspace{1cm}
\vspace*{-1cm} \caption{Diagrams considered for the calculation of 
N to $\Delta$ transitions that do not contain strange quark mass terms.}
 \label{xmag}
\end{figure}
%
%
\begin{figure}[htb]
\centerline{\epsfig{file=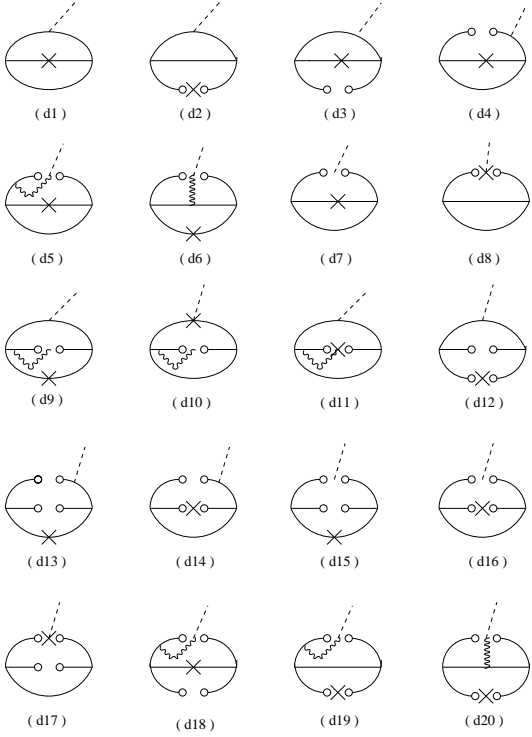,width=7.0cm}} \vspace{1cm}
\vspace*{-1.5cm} \caption{Diagrams considered for the calculation of 
N to $\Delta$ transitions that are proportional to the strange quark mass 
(denoted by $\times$).}
\label{xmagm}
\end{figure}

With the above elements in hand, it is straightforward to evaluate
the correlation function by substituting the quark propagator into
the various master formulas. We keep terms to first order in the
external field and in the strange quark mass (u-quark and d-quark masses are ignored). 
Vacuum condensates up to dimension eight are retained in the expansion. 
The algebra is extremely tedious. Each term in the master formula is 
a product of three copies of the quark propagator, and  
there are hundreds of such terms over various color permutations. 
The calculation can be organized by the diagrams
(similar to Feynmann diagrams) in Fig.~\ref{xmag} and
Fig.~\ref{xmagm}. Note that each diagram is only generic and all
possible color permutations are implied. We used the computer algebra 
package called REDUCE to carry out some of the calculations. 
The QCD side has the
same tensor structure as the phenomenological side and the results
can be collected according to the same 12 independent tensor structures in Eq.~(\ref{wewo}).


\subsection{The QCD sum rules} \label{qcdsr}

Once we have both the QCD side (commonly referred as LHS) and the phenomenological side
(RHS), we can isolate the sum rules by matching the two sides. Since
there are 12 independent tensor structures, 12 sum
rules can be constructed for each transition. 
For the 8 transitions, the total number of sum rules is 96. 
We divide the sum rules into four groups
according to their dependence on the amplitudes and performance.
The first group consists of the sum rules at
structures $\mbox{WE}_1$, $\mbox{WO}_2$ and $\mbox{WE}_3$ which 
only contain the amplitude $G_1$, and at $\mbox{WE}_6$ which only contains
$G_2$. These sum rules are expected to perform better because they involve fewer 
parameters.
The second group consists of the sum rules at structures $\mbox{WE}_5$,
$\mbox{WO}_5$ and $\mbox{WO}_6$ which have both $G_1$ and $G_2$.
It is difficult to extract both amplitudes at the same time.
One way out is to fix one of them and treat the other as free parameter.
This reduces the predictive ability of the QCD sum rules. Our numerical analysis suggests 
that it can serve as a consistence check at best.
Another way is to take appropriate combinations of sum rules to eliminate 
one of the amplitudes. We tried but found the sum rules obtained this way 
perform worse than the un-manipulated ones. We advocate against the practice of 
manipulating the original sum rules like algebraic equations 
(such as taking derivatives or linear combinations) to obtain new sum rules.
The third group consists of the sum rules at structures
$\mbox{WE}_2$, $\mbox{WO}_1$ and $\mbox{WO}_3$ which 
have one of the amplitudes and additional dependence on the averaged square mass $\bar m^2$, 
aside from the common factor $e^{-\bar m^2/M^2}$. 
We found these sum rules perform consistently worse than those in groups 1 and 2, 
possibly because of the approximation introduced by $\bar m^2$.
The fourth group consists of the sum rules at 
$\mbox{WE}_4$ and $\mbox{WO}_4$ which 
contain both $G_1$ and $G_2$, {\it and} additional $\bar m^2$ dependence.
They perform consistently the worst.
In the following, we are going to give detailed analysis
of the sum rules in group 1, but first let us present the sum rules 
in this group.

At the structure $\mbox{WE}_1$, the sum rules can be expressed in
the following form:
\begin{equation}
  \begin{array}{l}
   c_1 L^{-4/27} 
   E_2(w) M^{4} +c_2\chi a m_s L^{-20/27}  E_1(w) M^{2}
   \\+c_3 am_s L^{-4/27} E_0(w) +c_4 \chi a^2 L^{4/27}E_0(w)
   \\+  c_5b L^{-4/27}E_0(w)+c_6m_0^2 a m_s L^{-18/27} \frac{1}{M^2}
   \\+c_7  a^2 L^{20/27}\frac{1}{M^2}   + c_8  \chi m_0^2 a^2 L^{-10/27} \frac{1}{M^2}
\\ = \tilde \lambda _N \tilde \lambda _\Delta e^{-\bar m^2/M^2} [\frac{{ 1 }}{3M^2} G_1(m_{\Delta} +
m_N)+A].
\end{array}
\label {we1}
\end{equation}
Here $\tilde \lambda _N $ and $\tilde \lambda _\Delta $ are the rescaled
current coupling ${\tilde \lambda ^2 \equiv
(2\pi)^4\lambda^2}$. The quark condensate, gluon condensate, and
the mixed condensate are represented by
\begin{equation}
a=-(2\pi)^2\,\langle\bar{u}u\rangle, \hspace{2mm} b=\langle
g^2_c\, G^2\rangle, \hspace{2mm} m_0^2=-\langle\bar{u}g_c\sigma\cdot G
u\rangle/\langle\bar{u}u\rangle.
\end{equation}
The quark charge factors $e_q$ are given in units of electric charge
\begin{equation}
e_u=2/3, \hspace{4mm} e_d=-1/3, \hspace{4mm} e_s=-1/3.
\end{equation}
Note that we choose to keep the quark charge factors explicit in
the sum rules. The advantage is that it can facilitate the study
of individual quark contributions to the amplitudes. The
parameters $f$ and $\phi$ account for the flavor-symmetry breaking
of the strange quark in the condensates and susceptibilities:
\begin{equation}
f={ \langle\bar{s}s\rangle \over \langle\bar{u}u\rangle} ={
\langle\bar{s}g_c\sigma\cdot G s\rangle \over
   \langle\bar{u}g_c\sigma\cdot G u\rangle},
\hspace{4mm} \phi={ \chi_s \over \chi}={ \kappa_s \over \kappa}={
\xi_s \over \xi}.
\end{equation}
The anomalous dimension corrections of the interpolating fields
and the various operators are taken into account in the leading
logarithmic approximation via the factor
\begin{equation}
L^\gamma=[{\alpha_s(\mu^2) \over \alpha_s(M^2)}]^\gamma
=[{\ln(M^2/\Lambda_{QCD}^2) \over \ln(\mu^2/\Lambda_{QCD}^2)}
]^\gamma,
\end{equation}
where $\mu=500$ MeV is the renormalization scale and
$\Lambda_{QCD}$ is the QCD scale parameter. As usual, the pure
excited state contributions are modeled using terms on the OPE
side surviving $M^2\rightarrow \infty$ under the assumption of
duality, and are represented by the factors
\begin{equation}
E_n(w)=1-e^{-w^2/M^2}\sum_n{(w^2/M^2)^n \over n!},
\end{equation}
where $w$ is an effective continuum threshold and it is in
principle different for different sum rules and we will treat it
as a free parameter in the analysis.

The $c_i$ coefficients differ from transition to transition. It is not
necessary to list the coefficients separately for all 8 transition channels. We only
need to carry out three separate calculations for $\Sigma^+ \gamma
\to \Sigma^{*+}  $, $\Xi^0 \gamma \to \Xi^{*0}  $ and $\Lambda
\gamma \to \Sigma^{*0}$. Other channels can be obtained
from them by making appropriate substitutions as specified below:
    \begin{itemize}
    \item for $p \gamma \to \Delta^{+} $, replace s quark by d quark in $\Sigma^+ \gamma \to
\Sigma^{*+}  $,
    \item for $n \gamma \to \Delta^{0} $, interchange d quark and u quark in proton $p \gamma \to \Delta^{+} $,
    \item for $\Sigma^-   \gamma \to \Sigma^{*-}  $, replace u quark by d quark in $\Sigma^+   \gamma
\to \Sigma^{*+}  $,
     \item for $\Sigma^0 \gamma \to \Sigma^{*0}  $, add corresponding coefficients of $\Sigma^+   \gamma
\to \Sigma^{*+}  $ and $\Sigma^-   \gamma \to \Sigma^{*-}  $, then
divide by 2,
     \item for $\Xi^-
\gamma \to \Xi^{*0}  $, replace u quark by d quark in $\Xi^0
\gamma \to \Xi^{*0}  $.
    \end{itemize}
Here the interchange between u and d quarks is achieved by simply
switching their charge factors $e_u$ and $e_d$. The
conversions from s quark to u or d quarks involve setting $m_s=0$,
$f=\phi=1$, in addition to switching the charge factors. 
In practice, we computed all 8 channels separately, and used the 
substitutions as a check of our calculations.

The coefficients appearing in Eq.(~\ref{we1}) are given by, in the
transition channel $\Sigma^+ \gamma \to \Sigma^{*+}  $,
\begin{equation}
\begin{array}{l}
  c_1=\frac{1}{4} (-1+\beta) (e_s-e_u),
 \\c_2=-\frac{1}{4} (1+2 \beta) (e_u-e_s f_s  \phi ),
 \\c_3=\frac{1}{12} (e_u (-1+\beta+3  f_s+3\beta  f_s+\kappa-\beta \kappa-(2+\beta)\xi) \\ \;\;\; +e_s (-6+ f_s (4-\kappa  \phi +2 \phi  \xi)+\beta (-6+ f_s (2+ \phi (\kappa+\xi))))),
 \\c_4=\frac{1}{6} (-e_u(1+\beta+ f_s)+(2+\beta) e_s  f_s \phi ),
 \\c_5=-\frac{1}{48} (-1+\beta) (e_s-e_u),
 \\c_6=\frac{1}{24} (1+2 \beta) (2 e_s-e_u(1+ f_s)),
 \\c_7=\frac{1}{36} (-e_u (2-5\kappa+\xi+\beta(2+\kappa+\xi)+ f_s (-4+4\kappa+\xi))\\ \;\;\; +e_s (-6+ f_s (4+2\beta-\kappa  \phi +\beta \kappa \phi +(2+\beta)  \phi  \xi))),
 \\c_8=\frac{1}{144} (e_u (8+2 \beta+8  f_s+5\beta  f_s)-(16+7 \beta) e_s  f_s \phi
 );
\end{array}
\end{equation}

in the transition channel $\Xi^0\gamma \to \Xi^{*0} $,
\begin{equation}
\begin{array}{l}
  c_1=-\frac{1}{8} (-1+\beta) (e_s-e_u),
 \\c_2=\frac{1}{8} (1+3 \beta) (e_u-e_s f_s  \phi ),
 \\c_3=\frac{1}{12} (e_s  f_s-e_s  f_s \phi  (\kappa+\xi)+e_u (2-3 f_s+\kappa+\xi)\\ \;\;\;+\beta (e_u (8-3 f_s+\kappa+\xi)-e_s (6+ f_s(-1+ \phi  (\kappa+\xi))))),
 \\c_4=\frac{1}{6} (1+\beta)  f_s (e_u-e_s f_s  \phi ),
 \\c_5=\frac{1}{192} (-1+\beta) (e_s-e_u),
 \\c_6=\frac{1}{48} (e_s (2-\beta (-4+ f_s)-5 f_s)
  \\ \;\;\;+e_u (1+2  f_s+\beta (-7+4  f_s))),
 \\c_7=\frac{1}{36} (e_u (-3 (1+\beta) f_s^2+3 (1+\beta) \kappa \\ \;\;\;
  + f_s (2-2\kappa+\xi+\beta (8-2\kappa+\xi)))\\ \;\;\;+e_s (-3+ f_s (6-3\kappa  \phi + f_s (-2+2 \kappa \phi - \phi  \xi))
  \\ \;\;\;  +\beta (-3-3  f_s\kappa  \phi + f_s^2 (-2+2 \kappa \phi - \phi  \xi)))),
 \\c_8=-\frac{1}{144} (1+\beta) (e_u (-1+8 f_s)-e_s  f_s (2+5  f_s)  \phi
 );
  \end{array}
\end{equation}

and in the transition channel $\Lambda\gamma \to \Sigma^{*0} $,
\begin{equation}
 \begin{array}{l}
  c1=\frac{1}{8} (-(2+\beta) e_d+e_s+(2+\beta) e_u),
 \\c_2=\frac{1}{8} ((3-2 \beta) e_d+2 (-1+\beta) e_u),
 \\c_3=\frac{1}{24} (e_d (-2-12 f_s-\kappa+\beta (1+3
f_s-\kappa-\xi)+5 \xi)\\ \;\;\;+e_u (-2+6 f_s+5 \kappa-4 \xi+\beta
(-1-3 f_s+\kappa+\xi))\\ \;\;\;+e_s (-6+f_s (-2+2 \kappa \phi-\phi
\xi))),
 \\c_4=\frac{1}{12} (-e_u (2-3 \beta+2 (1+\beta) f_s)\\ \;\;\;+e_d (3+2
f_s+\beta (-3+2 f_s))-e_s f_s \phi),
 \\c_5=\frac{1}{192} ((5+2 \beta) e_d+3 e_s-(3+2 \beta) e_u),
 \\c_6=\frac{1}{96} (-2 e_s-(-9+4 \beta) e_d (1+f_s)+(-5+4 \beta) e_u
(1+f_s)),
 \\c_7=\frac{1}{72} (e_s f_s (-2+2 \kappa \phi-\phi \xi)+e_d (2 f_s
(-7-2 \kappa+\xi)\\ \;\;\;+3 (2+\kappa+\xi)+\beta (-3+2 f_s)
(2+\kappa+\xi))-e_u (2 (2+\kappa+\xi)\\ \;\;\;+\beta (-3+2 f_s)
(2+\kappa+\xi)+f_s (-8-7 \kappa+2 \xi))),
 \\c_8=\frac{1}{288} (-e_d (23+17 f_s+\beta (-20+13 f_s))\\ \;\;\;+e_u (15
(1+f_s)+\beta (-20+13 f_s))).
\end{array}
\end{equation}

At structure $\mbox{WE}_6$, the sum rule is:
\begin{equation}
  \begin{array}{l}
    c_1 
\chi a m_s L^{-20/27}  E_1(w) M^{2} +c_2 am_s L^{-4/27} E_0(w)
\\+c_3\chi a^2 L^{4/27}E_0(w) +c_4  a^2 L^{20/27} \frac{1}{M^2}
 +  c_5 \chi m_0^2 a^2 L^{-10/27} \frac{1}{M^2}
\\ = \tilde \lambda _N \tilde \lambda _\Delta e^{-\bar m^2/M^2}
[\frac{{ 1 }}{M^2} G_2(m_N ^2 -  m_{\Delta} m_N)+A].
 \end{array}
\label{we6}
\end{equation}
The coefficients appearing in Eq.(~\ref{we6}) are given by, in the
transition channel $\Sigma^+   \gamma \to \Sigma^{*+}  $,
\begin{equation}
\begin{array}{l}
  c_1=\beta (-e_u+e_s  f_s  \phi ),
 \\c_2=\frac{1}{3} \beta (e_u (-4-2\kappa+\xi)+e_s  f_s (4+2 \kappa \phi - \phi  \xi)),
 \\c_3=-\frac{2}{3} \beta (e_u-e_s  f_s \phi ),
 \\c_4=\frac{1}{18} (-3 e_u (-1+ f_s) \xi+2\beta (e_u (-4-2 \kappa+\xi)
   \\ \;\;\;+e_s f_s (4+2 \kappa  \phi - \phi  \xi))),
 \\c_5=\frac{1}{6} \beta (e_u-e_s  f_s \phi );
\end{array}
\end{equation}

in the transition channel $\Xi^0\gamma \to \Xi^{*0} $,
\begin{equation}
\begin{array}{l}
  c_1=\beta (e_u-e_s  f_s  \phi ),
 \\c_2=\frac{1}{3} \beta (e_u (4+2\kappa-\xi)+e_s  f_s (-4-2 \kappa \phi + \phi  \xi)),
 \\c_3=\frac{2}{3} \beta  f_s (e_u-e_s f_s  \phi ),
 \\c_4=\frac{1}{18}  f_s (3 e_s (-1+ f_s) \phi  \xi+2 \beta (e_u (4+2\kappa-\xi)\\ \;\;\;+e_s  f_s (-4-2 \kappa \phi + \phi  \xi))),
 \\c_5=\frac{1}{6} \beta  f_s (-e_u+e_s f_s  \phi );
  \end{array}
\end{equation}

and in the transition channel $\Lambda\gamma \to \Sigma^{*0} $,
\begin{equation}
\begin{array}{l}
  c_1=\frac{1}{2} \beta (e_d-e_u),
 \\c_2=\frac{1}{6} \beta (e_d-e_u) (4+2\kappa-\xi),
 \\c_3=\frac{1}{3} \beta (e_d-e_u),
 \\c_4=\frac{1}{36} (e_d-e_u) (\beta (8+4\kappa-2 \xi)-3 (-1+ f_s) \xi),
 \\c_5=\frac{1}{12} \beta (-e_d+e_u).
  \end{array}
\end{equation}

At structure $\mbox{WO}_2$, the sum rule is:
\begin{equation}
  \begin{array}{l}
    c_1 
m_s  L^{-16/27} E_1(w) M^{2} +c_2 \chi a L^{-8/27} E_1(w) M^{2}
\\+c_3 a L^{8/27} E_0(w) +c_4 m_0^2 a L^{-6/27} \frac{1}{M^2}   +
c_5
 \chi a^2 m_s L^{-8/27} \frac{1}{M^2} \\+c_6   \chi a b L^{-8/27} \frac{1}{M^2}
+c_7 a^2 m_s L^{8/27} \frac{1}{M^4}  +c_8  ab L^{8/27}
\frac{1}{M^4}
\\ = \tilde \lambda _N \tilde \lambda _\Delta e^{-\bar m^2/M^2}
[\frac{{ 1 }}{3M^2} G_1 \frac{1}{m_{\Delta}}(m_N + m_{\Delta})+A].
 \end{array}
\label{wo2}
\end{equation}

The coefficients appearing in Eq.(~\ref{wo2}) are given by, in the
transition channel $\Sigma^+   \gamma \to \Sigma^{*+}  $,
\begin{equation}
\begin{array}{l}
  c_1=\frac{1}{2} (-e_s+e_u),
 \\c_2=\frac{1}{6} (-e_u+e_s  f_s  \phi ),
 \\c_3=\frac{1}{12} (e_u (-2+6  f_s+2 \kappa+6\beta (-2+2  f_s+\kappa)-\xi)
   \\ \;\;\;+e_s(-6+ f_s (2-2 (1+3 \beta) \kappa  \phi + \phi \xi))),
 \\c_4=\frac{1}{12} (4 e_s-e_u (1+2 \beta(-1+ f_s)+3  f_s)),
 \\c_5=\frac{1}{6} (e_u (1+4 \beta (-1+ f_s)+3 f_s)-4 e_s  f_s  \phi ),
 \\c_6=\frac{1}{72} \beta (e_u-e_s  f_s \phi ),
 \\c_7=\frac{1}{36} (e_s (6+ f_s (-4+ \phi (\kappa+6 \beta \kappa-2 \xi))) \\ \;\;\;+e_u (2-2\kappa+\xi+ f_s (-4+\kappa-6 \beta\kappa+\xi))),
 \\c_8=\frac{1}{144} (-2 e_s+e_u (1-2 \beta(-1+ f_s)+ f_s));
\end{array}
\end{equation}

in the transition channel $\Xi^0\gamma \to \Xi^{*0} $,
\begin{equation}
\begin{array}{l}
  c_1=\frac{1}{4} (1+\beta) (e_s-e_u),
 \\c_2=-\frac{1}{12} (-1+\beta) (e_u-e_s f_s  \phi ),
 \\c_3=\frac{1}{24} (e_u (2-6 f_s+\kappa+\xi)-e_s  f_s (-4+ \phi (\kappa+\xi))
   \\ \;\;\;+\beta (e_u (-14+6  f_s-7\kappa-\xi)
    \\ \;\;\;+e_s (24+ f_s (-16+7 \kappa \phi + \phi  \xi)))),
 \\c_4=\frac{1}{48} (-2 (e_s+6  \beta e_s+e_u-6 \beta e_u)
   \\ \;\;\;+((-5+7 \beta)e_s+(9-7 \beta) e_u)  f_s),
 \\c_5=\frac{1}{12} (\beta (-1+ f_s) (2 e_u-7e_s  f_s  \phi )\\ \;\;\;+ f_s (-4 e_u+e_s(1+3  f_s)  \phi )),
 \\c_6=-\frac{1}{288} (-1+3 \beta) (e_u-e_s f_s  \phi ),
 \\c_7=\frac{1}{72} (e_s (12+2  f_s (-8-5\beta+ f_s+(2+\beta (-5+ f_s)- f_s)\\ \;\;\;\kappa  \phi )
  -(-1+\beta)  f_s (1+ f_s) \phi  \xi)+2 e_u (-3 \kappa+ f_s (-2+3 f_s\\ \;\;\;+2 \kappa-\xi)
    +\beta (2+\kappa+3 f_s (1+\kappa)+\xi))),
 \\c_8=\frac{1}{576} (-2 (-1+2 \beta)(e_s+e_u)\\ \;\;\;+(e_s+5 \beta e_s+(-5+3\beta) e_u)
 f_s);
  \end{array}
\end{equation}

and in the transition channel $\Lambda\gamma \to \Sigma^{*0} $,
\begin{equation}
 \begin{array}{l}
  c1=\frac{1}{4} (-3 e_d+e_s+2 e_u),
 \\c_2=\frac{1}{6}({e_d-e_u}),
 \\c_3=\frac{1}{24} (6 e_s+e_u (2+6 f_s+7 \kappa-6 \beta (-2+2
f_s+\kappa)-2 \xi)\\ \;\;\;+e_d (-8-6 f_s-7 \kappa+6 \beta (-2+2
f_s+\kappa)+2 \xi)),
 \\c_4=\frac{1}{96} (e_d (30+\beta (23-22 f_s)+16 f_s)\\ \;\;\;+e_u (-2 (7+8
f_s)+\beta (-23+26 f_s))),
 \\c_5=\frac{1}{6} (e_d (-3+2 \beta (-1+f_s)-2 f_s)\\ \;\;\;+2 e_u
(1+\beta+f_s-\beta f_s)+e_s f_s \phi),
 \\c_6=\frac{1}{288} ((-1+2 \beta) e_d-(1+2 \beta) e_u),
 \\c_7=\frac{1}{72} (-6 e_s-e_u (\kappa-12 \beta \kappa\\ \;\;\;+f_s (8+(4+6
\beta) \kappa-2 \xi)-2 (2+\xi))\\ \;\;\;+e_d (f_s (14+\kappa+6
\beta \kappa-2 \xi)-2 (2+(-2+6 \beta) \kappa+\xi))),
 \\c_8=\frac{1}{1152}(1-8 e_s+e_u (6+\beta (7-10 f_s)+8 f_s)\\ \;\;\;+e_d (-14-7 \beta-8
f_s+6 \beta f_s)).
\end{array}
\end{equation}

This concludes the presentation of the QCD sum rules that will be used 
in the numerical analysis.

%

\section{Sum Rule Analysis}
\label{ana}
\subsection{General procedure}
The sum rules for N to $\Delta$ transition have the generic form
of OPE - ESC = Pole + Transition, or
\begin{equation}
\begin{array}{l}
\Pi_{tran}(QCD,\beta,w,M^2) =
\\ \;\;\;\;\;\;\;\;\;\;\tilde{\lambda}_N
\tilde{\lambda}_{\Delta} \left({f(G_1,G_2)\over M^2} + A\right)
e^{-\bar m^2/M^2}, \end{array}\end{equation}
where $QCD$ represents all the QCD input parameters. $f(G_1,G_2)$
is a function of transition amplitudes $G_1$ and $G_2$. The basic mathematical task
is: given the function $\Pi_{tran}$ with known QCD input
parameters, 
find the phenomenological parameters ($G_1$, $G_2$, transition strength
$A$, coupling strength $\tilde{\lambda}$, and continuum threshold
$w$) by matching the two sides over some region in the Borel mass
$M$. A $\chi^2$ minimization is best suited for this purpose. It
turns out that there are too many fit parameters for this
procedure to be successful in general. To alleviate the situation,
we employ the corresponding mass sum rules which have a generic form of OPE - ESC = Pole, or
\begin{eqnarray}
\Pi_{mass, N}(QCD,\beta,w_1,M^2) = \tilde{\lambda}_N^2
e^{-m_N^2/M^2};\\
\Pi_{mass, \Delta}(QCD,\beta,w_2,M^2) = \tilde{\lambda}_\Delta^2
e^{-m_\Delta^2/M^2}.
\end{eqnarray}
They share some of the common parameters and factors with the transition sum rules. 
Note that the continuum thresholds may not be the same in different sum rules.
By taking the following combination of the transition and mass sum rules, 
\begin{equation}
\begin{array}{l}
\frac{\Pi_{tran}(QCD,\beta,w,M^2)}{
\sqrt{\Pi_{mass,N}(QCD,\beta,w_1,M^2) \Pi_{mass,
\Delta}(QCD,\beta,w_2,M^2)}} \\= \frac{f(G_1,G_2)}{M^2}+A,
\end{array}
\label{ratio}
\end{equation}
the couplings $\lambda$ and the exponential factors are canceled out.
This is another advantage for introducing the $\bar m^2$ prescription.
The form in Eq.~(\ref{ratio}) is what we are going to implement. By plotting the two
sides as a function of $1/M^2$, the slope will be related to the transition
amplitudes $G_1$ or $G_2$ and the intercept the transition
strength A. The linearity (or deviation from it) of the left-hand
side gives an indication of OPE convergence and the role of
excited states. The two sides are expected to match for a good sum
rule over a certain window. This way of matching the sum rules has two advantages.
First, the slope, which is proportional to the transition
amplitudes, is usually better determined than the intercept.
Second, by allowing the possibility of different continuum
thresholds, we ensure that both sum rules stay in their valid
regimes.

For the octet baryons, we use the chiral-even mass sum rules in
Ref.~\cite{Lee02} which are reproduced here in the same notation used in this work.
\begin{equation}
\begin{array}{l}
p_1  L^{-4/9} E_3 (w_1)M^6
                 + p_2 b  L^{-4/9} E_1 (w _1 )M^2
                + p_3 m_s a L^{4/9}
                \\+ p_4 a^2 L^{4/9}
                 + p_5 a^2 k_v L^{4/9}
                + p_6 m_0^2 a^2 L^{-2/27}\frac{1}{M^2}\\=\tilde \lambda _N ^2 e^{ - m_N ^2 /M^2
               }.
\end{array}
\label{omass}
 \end{equation}
The coefficients are, for N:
\begin{equation}
\begin{array}{*{20}l}
   \begin{array}{l}
   p_1=\frac{1}{64} (5 +2 \beta  + 5\beta ^2),
   \;\;\;p_2= \frac{1}{256} (5 +2 \beta  + 5\beta ^2),
   \\p_3=0,
   p_4= \frac{1}{24}(7 -2\beta  -5\beta ), \;\;\;
   p_5=0,\;\;\;\\
   p_6=  - \frac{1}{96}(13 - 2\beta  - 11\beta ^2 );
\end{array}
\end{array}
\label{massp}
 \end{equation}
for $\Lambda$:
\begin{equation}
\begin{array}{*{20}l}
   \begin{array}{l}
p_1=\frac{1}{64} (5+2 \beta+5 \beta^2),
 \\ p_2= \frac{1}{256} (5+2 \beta+5 \beta^2),
 \\p_3=\frac{1}{96}((20-15f_s)-(16+6f_s)\beta-(4+15 f_s)\beta^2),
 \\ p_4=\frac{1}{96}((4f_s-5-6t)+(4+4f_s)\beta+(4 f_s+1+6
 t)\beta^2),
 \\ p_5= \frac{1}{72} ((10 f_s+11)+(2-8 f_s) \beta-(2 f_s+13)
 \beta^2),
 \\ p_6= \frac{1}{288} ((-16 f_s-23)+(8 f_s-2) \beta+(8 f_s+25)
 \beta^2);
 \end{array}
\end{array}
\label{masslam}
 \end{equation}
for $\Sigma$:
\begin{equation}
\begin{array}{*{20}l}
   \begin{array}{l}
p_1=\frac{1}{64} (5+2 \beta+5 \beta^2),
 \\ p_2=\frac{1}{256}(5+2 \beta+5 \beta^2),
 \\ p_3=\frac{1}{32}((12-5 f_s)-2 f_s \beta-(12+5f_s) \beta^2),
\\ p_4=- \frac{1}{94}((4 f_s+21+18 t)+4 f_s \beta +(4 f_s-21-18t)
\beta^2),
 \\ p_5=\frac{1}{24} ((6 f_s+1)-2 \beta-(6 f_s-1) \beta^2),
 \\ p_6=-\frac{1}{96}((12 f_s+1)-2 \beta-(12 f_s-1)\beta^2);
 \end{array}
\end{array}
\label{masssig}
 \end{equation}
for $\Xi$:
\begin{equation}
\begin{array}{*{20}l}
   \begin{array}{l}
p_1=  \frac{1}{64} (5+2 \beta+5 \beta^2),
 \\ p_2= \frac{1}{256} (5+2 \beta+5 \beta^2),
 \\ p_3= \frac{3}{16} ((2-f_s)-2 f_s \beta-(2+f_s) \beta^2),
\\ p_4=-\frac{1}{96}((15-f_s+18 t)-10 f_s \beta-(15+f_s+18 t)
\beta^2),
 \\ p_5= \frac{1}{24} f_s ((f_s+6)-2 f_s \beta+(f_s-6) \beta^2),
 \\ p_6= - \frac{1}{94} f_s ((f_s+12)-2 f_s \beta+(f_s-12) \beta^2).
\end{array}
\end{array}\label{massxi}
 \end{equation}
The function t is defined as $t\equiv \ln{\frac{M^2}{\mu^2}}-\gamma_{EM}$
with $\gamma_{EM}\approx 0.577$ the Euler-Mascheroni constant.

For the decuplet baryons, we use the chiral-odd sum
rules~\cite{Lee98} at the structure $g_{\scriptscriptstyle
\mu\nu}$, which are reproduced here, for $\Delta$:
\begin{equation}
\begin{array}{l}
{4\over 3}\,a\; E_1\,L^{16/27}\;M^4 - {2\over 3}\,m^2_0 a\;
E_0\,L^{2/27}\;M^2 - {1\over 18}\,a\,b\; L^{16/27}
\\=\tilde{\lambda}_{\scriptscriptstyle \Delta}^2\;
M_{\scriptscriptstyle \Delta}\;e^{-M_{\scriptscriptstyle
\Delta}^2/M^2};
\label{Delta1}
\end{array}
\end{equation}
for ${\Sigma^*}$:
\begin{equation}
\begin{array}{l}
{4\over 9}\,(f_s+2)\,a\; E_1\,L^{16/27}\;M^4 \\-{2\over
9}\,(f_s+2)\,m^2_0 a\; E_0\,L^{2/27}\;M^2 - {1\over
54}\,(f_s+2)\,a\,b\; L^{16/27} \nonumber \\ +{1\over 2}\,m_s\;
E_2\,L^{-8/27}\;M^6 + {2\over 3}\,m_s\,\kappa_v a^2\;L^{16/27}
\\=\tilde{\lambda}_{\scriptscriptstyle {\Sigma^*}}^2\;
M_{\scriptscriptstyle {\Sigma^*}}\;e^{-M_{\scriptscriptstyle
{\Sigma^*}}^2/M^2}; \label{Sigma1}
 \end{array}
 \end{equation}
for ${\Xi^*}$:
\begin{equation}
\begin{array}{l}
{4\over 9}\,(2f_s+1)\,a\; E_1\,L^{16/27}\;M^4
\\-{2\over
9}\,(2f_s+1)\,m^2_0 a\; E_0\,L^{2/27}\;M^2 - {1\over
54}\,(2f_s+1)\,a\,b\; L^{16/27} \nonumber \\  +m_s\;
E_2\,L^{-8/27}\;M^6 + {4\over 3}\,m_s\,f_s\,\kappa_v
a^2\;L^{16/27}
\\=\tilde{\lambda}_{\scriptscriptstyle
{\Xi^*}}^2\; M_{\scriptscriptstyle
{\Xi^*}}\;e^{-M_{\scriptscriptstyle {\Xi^*}}^2/M^2}. \label{Xi1}
\end{array}\end{equation}

\subsection{QCD input parameters}
The standard vacuum condensates are
taken as $a=0.52$ GeV$^3$, $b=1.2$ GeV$^4$, $m^2_0=0.72$ GeV$^2$. 
We take into account the possible factorization violation in the 
four-quark condensate (in terms of $\kappa_v a^2$) and use $\kappa_v=2.0$. 
The strange quark parameters are placed at $m_s=0.15$ GeV, $f=0.83$,
$\phi=0.60$~\cite{Pasupathy86,Lee98b,Lee98c}. 
The QCD scale parameter is restricted to $\Lambda_{QCD}=0.15$ GeV. 
This set of parameters is used in all studies of hadron properties in the QCD sum rule 
approach.
In the presence of external fields, 
additional condensates called vacuum susceptibilities are introduced.
These parameters are less well-known. 
They have been estimated in studies of nucleon magnetic
moments~\cite{Ioffe84,Chiu86,Wang08,MAM} where the availability of precise 
experimental data has put strict constraints on these parameters. 
In particular, the magnetic susceptibility $\chi$ is the subject of two recent
studies~\cite{Ball03,Rohrwild07a}.
We use the values $\chi=-6.0\; GeV^{-2}$ and $\kappa=0.75$, $\xi=-1.5.$
Note that $\chi$ is almost an order of magnitude larger than
$\kappa$ and $\xi$, and is the most important of the three. 

The above parameters are just central values. We will explore sensitivity to these
parameters by assigning uncertainties to them.
To this end, we use the Monte-Carlo procedure first introduced in Ref.~\cite{Derek96} 
which allows the most rigorous error analysis of QCD rum rules.
The procedure explores the entire phase-space of the input QCD parameters
simultaneously, and maps it into uncertainties in the phenomenological parameters.
It goes briefly as follows.
First, a sample of randomly-selected, Gaussianly-distributed condensates
are generated with assigned uncertainties. Here we give 10\% for
the uncertainties of input parameters, and this number can be
adjusted to test the sensitivity to the QCD parameters.  Then the OPE
is constructed in the Borel window with evenly distributed points $M_j$.
Note that the uncertainties in the OPE are not uniform
throughout the Borel window. They are larger at the lower end where
uncertainties in the higher-dimensional condensates dominate. Thus,
it is crucial that the appropriate weight is used in the calculation
of $\chi^2$. For the OPE obtained from the k'th set of QCD
parameters, the $\chi^2$ per degree of freedom is
 \begin{equation}
 {\chi^2_k\over N_{DF}}= \sum^{n_{\scriptscriptstyle
 B}}_{j=1} { [\Pi^{\scriptscriptstyle OPE}_k(M^2_j)
 -\Pi^{\scriptscriptstyle Phen}_k(M^2_j)]^2 \over
 (n_B-n_p)\;\sigma^2_{\scriptscriptstyle OPE}(M_j)},
 \end{equation}
where $\Pi^{\scriptscriptstyle OPE}$ refers to the LHS of Eq.~(\ref{ratio}) and
 $\Pi^{\scriptscriptstyle Phen}$ its RHS.
 The integer $n_p$ is the number of phenomenological search parameters.
 In this work, $n_B$=51 points were used along the Borel window. 
The procedure is repeated for many QCD parameter samples,
resulting in distributions for phenomenological fit parameters, from
which errors are derived. In practice, 200 samples are
sufficient for getting stable uncertainties. We used about 2000 samples
to resolve more subtle correlations among the QCD parameters
and the phenomenological fit parameters.
This means that each sum rule is fitted 2000 times to arrive at the final results.

\section{Result and discussion}
\label{res}
We did a wide survey of the 96 sum rules at the 12 structures and the 8 transitions.
Overall, based on the quality of the match,
the broadness of the Borel window and its reach into the lower
end, the size of the continuum and non-diagonal contributions, and OPE
convergence, we found that most of the sum rules do not perform well in terms of their 
ability to predict stable values for the amplitudes.
They can be roughly divided into four groups using our naming convention of 
$\mbox{WE}_i$ and $\mbox{WO}_i$, as alluded to earlier. 
Group one consistently produces the most reliable 
results, except the structure at $\mbox{WE}_3$. 
In the following, we focus on the sum rules at $\mbox{WE}_1$ and $\mbox{WO}_2$, 
which only contain the amplitude $G_1$, and at $\mbox{WE}_6$ which only contains $G_2$. 

\subsection{The sum rules at $\mbox{WE}_1$}
The sum rules are found in Eq.~(\ref{we1}). 
Ideally, we would like to extract 6 parameters: $G_1$, $A$, $w$, $w_1$,$w_2$
and $\beta$. But a search treating all six parameters as free does
not work because there is not enough information in the OPE. In
fact, the freedom to vary $\beta$ can be used as an advantage to
yield the optimal match. One choice is $\beta=-0.2$
which minimizes the perturbative term in the mass sum rule (the first term 
in Eq.~(\ref{omass}))~\cite{Narison}. Another parameter
that can be used to our advantage is the continuum thresholds $w_1$ and $w_2$
for the corresponding mass sum rules. We fix them to the values that
give the best solution to the mass sum rules independently. The
following values are used: for the nucleon,
$w_1=1.44$ GeV; for $\Lambda$, $w_1=1.60$ GeV; for $\Sigma$,
$w_1=1.66$ GeV; for $\Xi$, $w_1=1.82$ GeV;  for the $\Delta$,
$w_2=1.65$ GeV; for $\Sigma^*$, $w_2=1.80$ GeV; for $\Xi^*$,
$w_2=2.0$ GeV.  In this way the transition and the
mass sum rules can stay in their respective valid Borel regimes.
This leaves us with three parameters: $G_1$, $A$, $w$.
Unfortunately, a three-parameter search is either unstable or
returns values for $w$ smaller than the particle masses, a clearly
unphysical situation. Again we think this is a symptom of
insufficient information in the OPE to resolve the parameters. 
So we are forced to fix the continuum threshold $w$ to the value that corresponds 
to the best match for the central values of the QCD parameters, and 
extract two parameters: $G_1$ and $A$.

\begin{table}[thb] 
\caption{Results for $G_1$ in $N\to\Delta$ transitions from
the QCD sum rule at structure $\mbox{WE}_1$ in Eq.~(\protect\ref{we1}).
The columns correspond to, from left to
right: transition channel, $\beta$ value, Borel window, continuum threshold, 
transition strength $A$, amplitude $G_1$.
The errors in the results are derived from
2000 samples in the Monte-Carlo analysis with 10\% uncertainty on
all the QCD input parameters.} \label{tabwe1}
\begin{tabular}{cccccc}
\hline\hline   Transition & $\beta$  & Window & $w$  & A &   $G_1$  \\

 &  & (GeV) & (GeV) & (GeV$^{-2}$)  & (GeV$^{-1}$)  \\

\hline   $p \gamma \to \Delta^{+}$ & -0.2 & 0.95 to 1.3 &1.4 & -0.54(7) & 3.84 (37)  \\

$n \gamma \to \Delta^{0}$  & -0.2 & 0.95 to 1.3 & 1.4 & -0.54(7) & -3.84 (37) \\

$\Sigma^{+} \gamma \to \Sigma^{*+}$  & -0.2 & 1.2 to 1.4& 1.7  & 0.07(1) & 2.74 (25) \\

$\Sigma^{0} \gamma \to \Sigma^{*0}$ & -0.2 & 1.1 to 1.6 & 1.6 & -0.02(1) & 1.18 (10) \\

$\Sigma^{-} \gamma \to \Sigma^{*-}$ & -0.2 & 1.2 to 1.8 &1.9 & 0.01(1)& -0.33 (5)  \\

$\Xi^0 \gamma \to \Xi^{*0}$ & -0.2 & 1.2 to 1.8 & 1.8 & 0.04(1) & -1.02(10) \\

$\Xi^- \gamma \to \Xi^{*-}$ & -0.2 & 1.2 to 2.0 & 2.1 & 0(1) & 0.10 (2) \\

$\Lambda \gamma \to \Sigma^{*0}$  & -0.2 & 1.0 to 1.4& 1.65 & -0.01(1) & 2.92(25) \\

\hline\hline \end{tabular}
\end{table}
%

\begin{figure}[htb]
    \begin{center}
      \epsfig{file=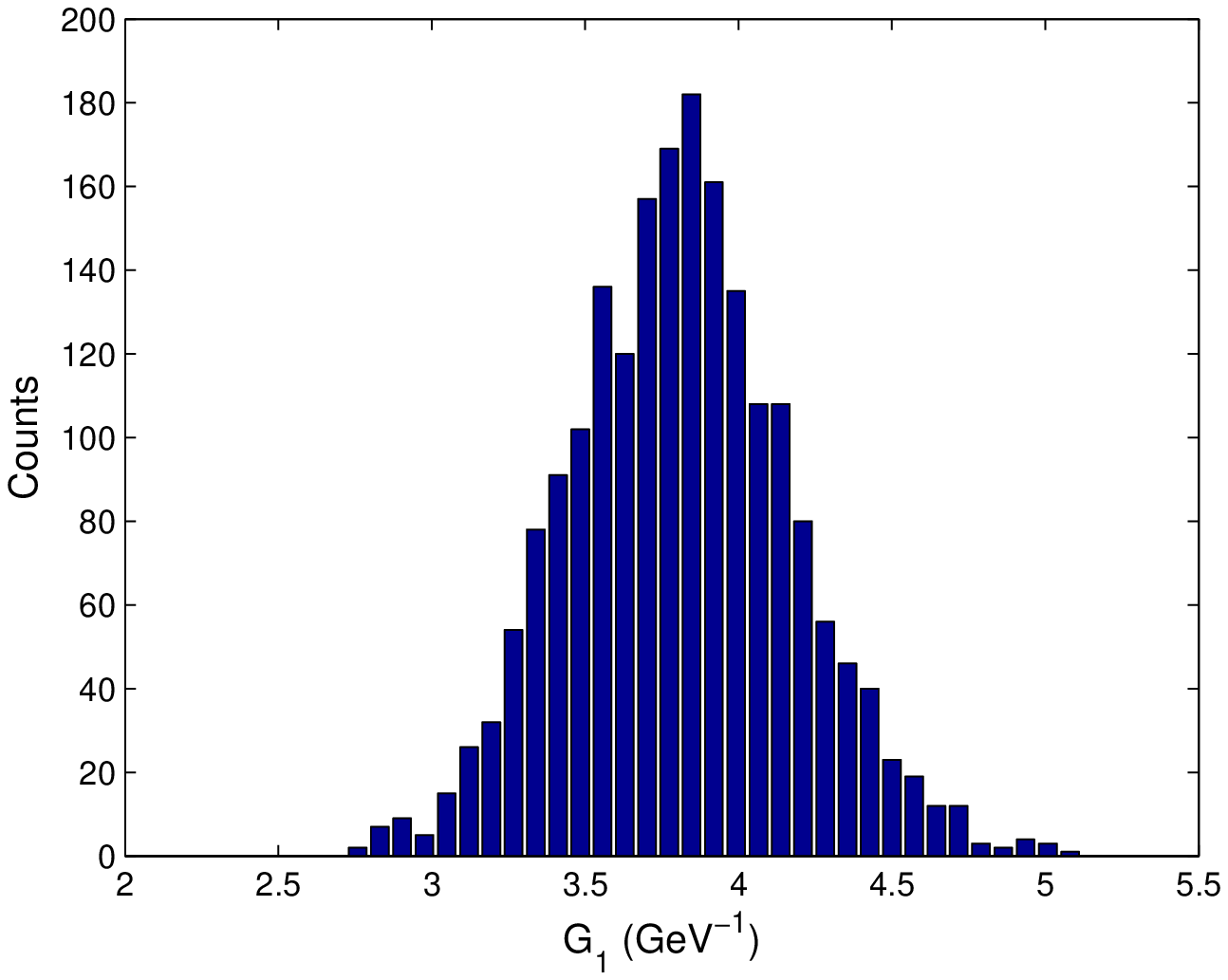, scale=0.50}
      \end{center}
\vspace{-0.6cm}
    \begin{center}
      \epsfig{file=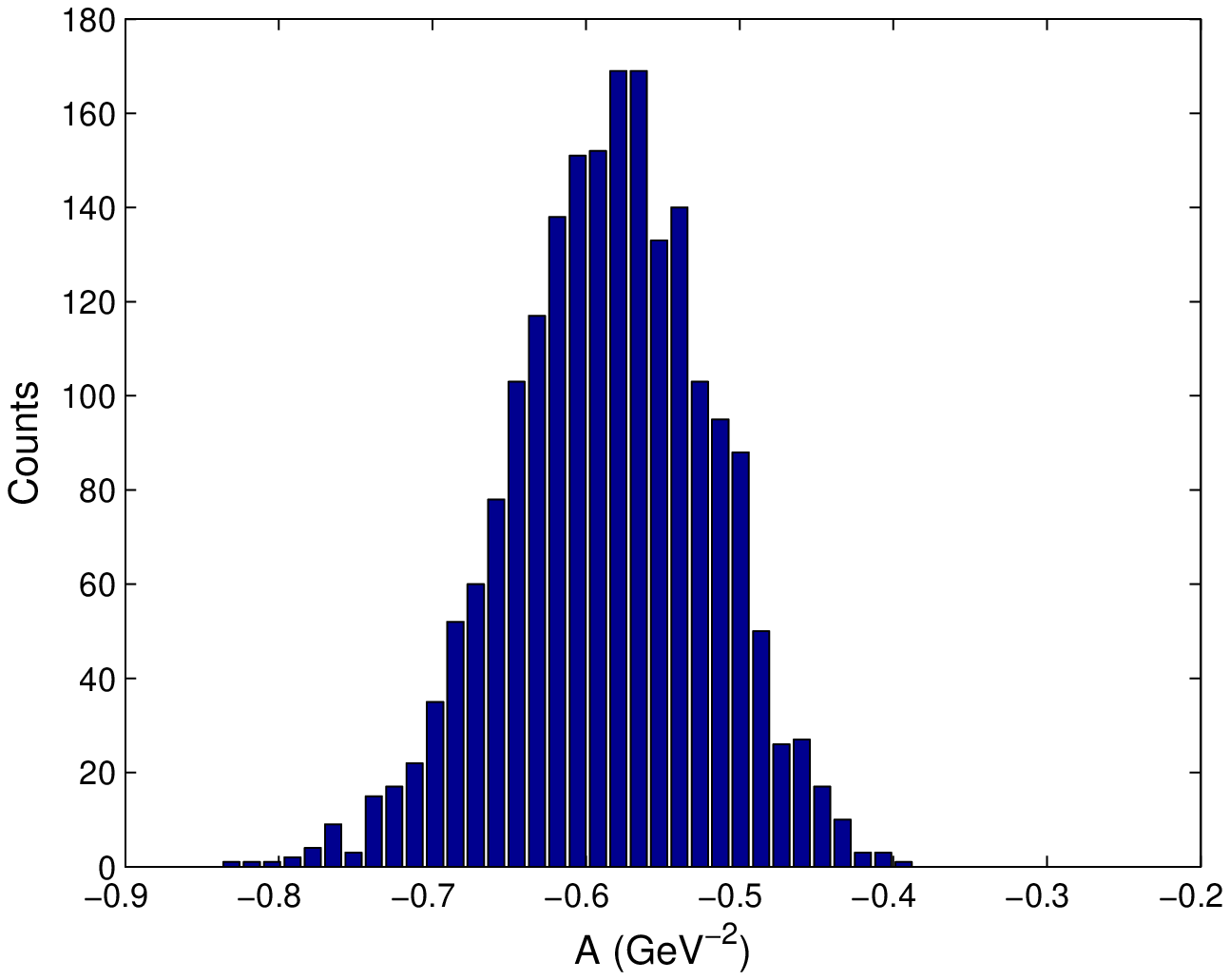, scale=0.50}
      \end{center}
\vspace{-0.8cm}
  \caption{Histogram for the $G_1$ transition amplitude (top) and
transition amplitude (bottom) obtained from Monte-Carlo fits of
Eq.(~\protect\ref{we1}) at $\mbox{WE}_1$ for 2000 QCD parameter
sets. They are based on 10\% uncertainty given to all the QCD
input parameters.}
  \label{histo}
\end{figure}

The results of such an analysis are given in Table~\ref{tabwe1}.
The Borel window is determined by the following two criteria: 
OPE convergence which gives the lower
bound, and ground-state dominance which gives the upper bound. It
is done iteratively: using the optimal value of $\beta$, we adjust
the Borel window until the best solution is found and the two criteria 
are roughly satisfied.
We also checked that the results are not
sensitive to small changes in $\beta$ and the Borel window.
Our result on the $\Lambda \gamma \to \Sigma^{*0}$ transition 
is a prediction: we know of no other calculations of this transition channel.
The results on the parameter $A$ indicate that the non-diagonal transitions 
are not significant in this sum rule, perhaps due to cancellations in the excited states, 
but they are not negligible in the proton and neutron channels. 
Ignoring them will alter the slope of the RHS and lead to different results for $G_1$.
In fact, we found that one of the main symptoms of a sum rule performing poorly is the 
relatively large contribution of the non-diagonal transitions represented by $A$.
The effects of such contributions are not known until a numerical analysis is carried out.
We stress that the errors are derived from Monte-Carlo
distributions which give the most realistic estimation of the
uncertainties. An example of such distributions is given in
Fig.~\ref{histo}. We see that they are roughly Gaussian
distributions. The central value is taken as the average, and the
error is one standard deviation of the distribution. We found
about 10\% accuracy in our Monte-Carlo analysis, 
resulting from 10\% uniform uncertainty in all the QCD
input parameters. Of course, the uncertainties in the QCD
parameters can be non-uniform. For example, we tried the
uncertainty assignments (which are quite conservative) in
Ref.~\cite{Derek96}, and found about 30\% uncertainties in our output.

%

\begin{figure}[tbh]
    \begin{center}
      \epsfig{file=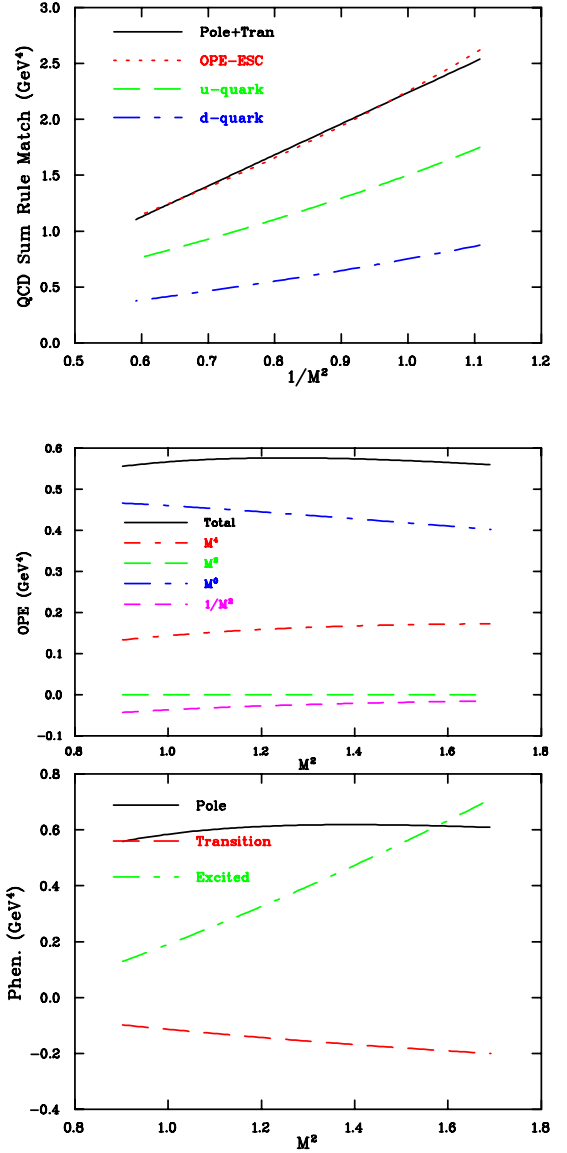, scale=0.68}
      \caption{Analysis of the QCD sum rule in Eq.~(\protect\ref{we1})
(structure $\mbox{WE}_1$) for the proton at $\beta=-0.2$ according
to Eq.~(\protect\ref{ratio}). Top figure, the pole plus transition
terms (solid lines) are compared against the OPE minus the
excited-state contributions (dashed lines) as a function of
$1/M^2$ (the two should match for an ideal sum rule). Also plotted
are the individual contributions from u (long-dashed lines) and d
(dot-dashed lines) quarks. Middle figure, the total in the OPE
side and its various terms are plotted as a function of $M^2$.
Bottom figure, the 3 terms in the phenomenological side: pole
(solid), transition (long-dashed), and excited (dot-dashed) are
plotted as a function of $M^2$.}
      \label{pwe1}
    \end{center}
\end{figure}
%

To gain a better appreciation on how the QCD sum rules produce the
results, we show some details of the analysis in Fig.~\ref{pwe1}, 
using the transition $p \gamma \to \Delta^{+}$ as an example.
There are three graphs in this figure to give three different
aspects of the analysis. The first graph shows how the two sides
of Eq.~(\protect\ref{ratio}) for this sum rule match over the Borel window, which
should be linear as a function of $1/M^2$ according to the
right-hand side of this equation. Indeed, we observe good
linear behavior from the LHS (which is OPE-ESC). 
The slope is directly proportional to the transition amplitude $G_1$, 
and the intercept give the non-diagonal transition contribution $A$.
We also plot the individual
contributions from u and d quarks. We see that
in this transition, the u-quark contribution is the dominant one,
which is expected because it is doubled-represented in the proton ($uud$).
In the next two graphs, we give details on the individual terms 
in the sum rule in Eq.~(\ref{we1}), remembering that the sum rule 
is in the generic form of OPE-ESC=Pole+Transition.
The second graph in Fig.~\ref{pwe1} shows how the various terms 
contribute to the OPE as a function of $M^2$. 
The $M^0$ terms, which contain the contributions from the condensates $\chi
a^2$ and $b$, play an important role. It is followed by the $M^4$ term which 
is the perturbative contribution. The $M^2$ term is zero in this channel because 
it is proportional to the strange quark mass. 
The $M^{-2}$ term gives a negative contribution and its contribution as a percentage 
of the entire OPE is about 10\% at the lower end of the Borel window.
In the third graph we show the
three terms that comprise the phenomenological side (Pole, Transition, and
ESC) as a function of $M^2$. The ground-state pole is dominant
at the low end of the Borel window. The excited-state contribution
starts small, then grows with $M^2$, as expected from the
continuum model. The transition contribution is small in this sum
rule. It is consistently smaller than the excited-state
contribution and has a weak dependence on the Borel mass.

\begin{figure}[tbh]
     \begin{center}
       \epsfig{file=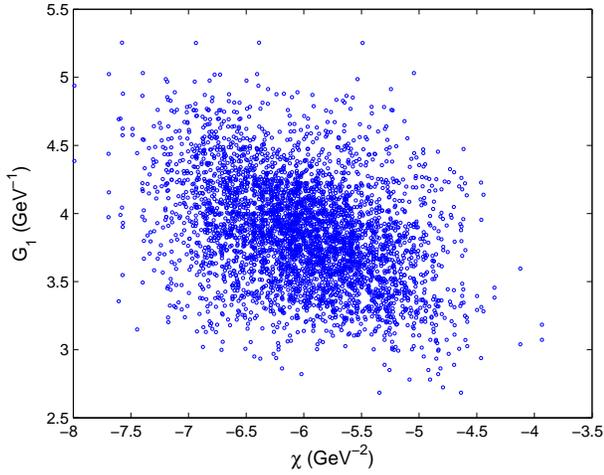, scale=0.60}
       \end{center}
   \caption{A scatter plot showing correlations between the transition amplitude $G_1$
 and one of the QCD parameters $\chi$ for the $p\gamma\to\Delta^+$ at structure
 $\mbox{WE}_1$. It is obtained from 2000 Monte-Carlo samples
 with 10\% uncertainty on all the QCD parameters.}
   \label{scapwe1}
 \end{figure}

In our Monte-Carlo analysis, the entire QCD input
phase space is mapped into the phenomenological output space,  
so we can also look into correlations between any two parameters by
way of scatter plots of the two parameters of interest. Fig.~\ref{scapwe1} 
shows such an example. We see that the transition amplitude $G_1$ 
has a strong negative correlation
with the vacuum susceptibility $\chi$. Larger $\chi$ (in absolute
terms since $\chi$ is negative) leads to smaller $G_1$. Precise
determination of the QCD parameters, especially  those that
have strong correlations to the output parameters, is crucial for
keeping the uncertainties in the spectral parameters under
control. We found similar strong correlations with $\chi$ in other transition channels. 

\subsection{The sum rule at $\mbox{WO}_2$}
The sum rule at this structure is found in Eq.~(\ref{wo2}).
We use the same procedure to analyze it. 
Since OPE expansion in this sum rule goes deeper than others (from $M^2$ to $1/M^4$),
it is expected to be more
reliable than the $\mbox{WO}_1$ sum rule. But our analysis shows
that this advantage is offset by the smallness of the $1/M^2$ and
$1/M^4$ terms. As a result, its performance is about the same as the sum rule at $\mbox{WE}_1$.
Table~\ref{tabwo2} displays the results extracted from this sum rule.
%

%
%
\begin{table}[thb] 
\caption{Similar to Table~\protect\ref{tabwe1}, but for the QCD
sum rule in Eq.~(\protect\ref{wo2}) (structure $\mbox{WO}_2$).}
\label{tabwo2}
\begin{tabular}{cccccc}
\hline\hline   Transition & $\beta$  & Window & $w$ &  A &   $G_1$  \\
 &  & (GeV) & (GeV) & (GeV$^{-3}$)  & (GeV$^{-1}$)  \\
\hline   $p \gamma \to \Delta^{+}$ & -0.2 & 0.9 to 1.2 & 1.3 & -0.5(1) & 3.52(32) \\

$n \gamma \to \Delta^{0}$  & -0.2 & 0.9 to 1.2 & 1.35 & 0.5(1)& -3.68(33) \\

$\Sigma^{+} \gamma \to \Sigma^{*+}$  & -0.2 & 1.2 to 1.6 & 1.4 & 0(1) & 2.20(20) \\

$\Sigma^{0} \gamma \to \Sigma^{*0}$ & -0.2 & 1.0 to 1.8 & 1.6 & 0.02(1) & 0.92(10) \\

$\Sigma^{-} \gamma \to \Sigma^{*-}$ & -0.2 & 1.0 to 1.6 & 1.6 & -0.01(1) & -0.32(5) \\

$\Xi^0 \gamma \to \Xi^{*0}$ & -0.2 & 1.0 to 1.8 & 1.65 & 0.01(1) & -0.91(10) \\

$\Xi^- \gamma \to \Xi^{*-}$ & -0.2 & 1.2 to 1.8 & 1.8 & 0(1) & 0.21(3)  \\

$\Lambda \gamma \to \Sigma^{*0}$  & -0.2 & 1.0 to 1.6 & 1.45 & 0.05(1) & 0.89(10) \\

\hline\hline \end{tabular}
\end{table}
%

\subsection{The sum rule at $\mbox{WE}_6$}
The sum rule is found in Eq.~(\ref{we6}). This is the only sum rule that 
depends on $G_2$ alone.
The results of our analysis are given in Table~\ref{tabwe6}.
The off-diagonal contributions (A) in the $p\to\Delta^+$ and $n\to\Delta^0$  
are smaller by half than those for $G_1$, but the uncertainties are larger. 
They are small in the other channels, just like those for $G_1$.
rediction, but others are not good. The reason of this is because
he transition contribution ($A$) are as larger as the $G_2$ term,
so the two terms can't distinguish and get contaminated.

%
\begin{table}[hbt] 
\caption{Similar to Table~\protect\ref{tabwe6}, but for the $G_{2}$ amplitude 
from the QCD sum rule in Eq.~(\protect\ref{we6}) (structure $\mbox{WE}_6$).}
\label{tabwe6}
\begin{tabular}{cccccc}
\hline\hline   Transition & $\beta$  & Window & $w$  & A &   $G_2$ \\
&  & (GeV) & (GeV) & (GeV$^{-2}$)  & (GeV$^{-2}$)  \\
\hline   $p \gamma \to \Delta^{+}$ & -0.2 & 1.0 to 1.3  & 1.3 & -0.26(4) &
-2.34(48)  \\
$n \gamma \to \Delta^{0}$  & -0.2 & 1.0 to 1.3  & 1.3 & 0.26(4) & 2.34(48)
 \\
$\Sigma^{+} \gamma \to \Sigma^{*+}$  & -0.2 & 1.4 to 1.6 & 1.8 & -0.07(1) &
-0.41(45)  \\
$\Sigma^{0} \gamma \to \Sigma^{*0}$ & -0.2 & 1.5 to 1.8 & 1.5 & -0.01(1) &
-0.31(16)  \\
$\Sigma^{-} \gamma \to \Sigma^{*-}$ & -0.2 & 1.2 to 1.6 & 1.8 & 0.01(1) &
-0.32(12) \\
$\Xi^0 \gamma \to \Xi^{*0}$ & -0.2 & 1.5 to 1.8 & 1.8 & 0.03(1) & 0.38(29)  \\
$\Xi^- \gamma \to \Xi^{*-}$ & -0.2 & 1.4 to 1.8 & 1.9 & 0(1) & 0.18(7)  \\
$\Lambda \gamma \to \Sigma^{*0}$  & -0.2 & 1.0 to 1.3 & 1.45 & -0.09(1)
&-0.62(17)  \\
\hline\hline \end{tabular}
\end{table}
%

\subsection{Individual quark contributions}
To gain a deeper understanding of the dynamics, we 
consider the individual quark sector contributions to the
transition amplitudes. In our approach, we can easily dial
individual quark contributions to the QCD sum rules. For example,
to turn off all u-quark (d-quark) contributions, we set the charge factor
$e_u=0$ ($e_d=0$). To turn off all s-quark contributions, we set $e_s=0$,
$m_s=0$, $f=1$, and $\phi=1$. We can extract a number
corresponding to each quark contribution from the slope of
Eq.~(\ref{ratio}) as a function of $1/M^2$, while keeping 
other factors the same ($\beta$, Borel window, $w$). We call this the raw
individual quark contributions to the transition amplitudes.
Table~\ref{g1uds} gives the results for the $G_{1}$ amplitude from such a study.
We see that the u-quark contribution in the transition $p \gamma \to \Delta^{+}$ 
is twice the d-quark contribution as expected, and both contributions are positive. 
It is the opposite in the transition $n \gamma \to \Delta^{0}$.
The s-quark contribution is positive in the $\Sigma$ transitions, and negative in the 
$\Xi$ and $\Lambda$ transitions. 
In the $\Sigma^{0} \gamma \to \Sigma^{*0}$ channel, 
the d-quark and s-quark contributions largely cancel.

Table~\ref{g1uds} gives the results for the $G_{2}$ amplitude.
Similar pattern is observed in the proton and neutron channels, albeit the signs are 
opposite for $G_1$ and $G_2$.
The s-quark contribution is opposite to that in $G_1$: 
negative in the $\Sigma$ channels and positive in the $\Xi$ channels. 
It donimates in the $\Sigma$ channels over the u-quark and d-quark contributions.
Interestingly, s-quark contribution is exactly zero in the $\Lambda$ channel.
%
\begin{table}
\caption{Individual quark contributions to the transition
amplitude $G_1$ in units of GeV$^{-1}$ extracted from the QCD sum rules 
in Eq.~(\protect\ref{we1}) (structure $\mbox{WE}_1$).} 
\label{g1uds}
\begin{tabular}{ccccccccc}
\hline\hline
 & $G_{1}^{u}$  & $G_{1}^{d}$ & $G_{1}^{s}$   & $G_{1}^{tot}$ & \\
\hline
$p \gamma \to \Delta^{+}$ & 2.56(25)& 1.28(13 )& 0& 3.84(38) \\
$n \gamma \to \Delta^{0}$& -1.28(13)& -2.56(25)& 0& -3.84(38) \\
$\Sigma^{+} \gamma \to \Sigma^{*+}$& 2.06(17)& 0& 0.69(6)& 2.75(22) \\
$\Sigma^{0} \gamma \to \Sigma^{*0}$& 1.00(9)& -0.50(5)& 0.67(6)& 1.17(10) \\
$\Sigma^{-} \gamma \to \Sigma^{*-}$& 0& -1.04(8)& 0.71(6)& -0.33(6) \\
$\Xi^0 \gamma \to \Xi^{*0}$& -0.71(6)& 0& -0.25(2)& -0.97(7) \\
$\Xi^- \gamma \to \Xi^{*-}$& 0& 0.37(3)& -0.27(2)& 0.11(2) \\
$\Lambda \gamma \to \Sigma^{*0}$& 1.91(16)& 1.17(10)& -0.16(2)& 2.92(25) \\
\hline\hline
\end{tabular}
\end{table}
%
\begin{table}[hbt] 
\caption{Individual quark contributions to the transition
amplitude $G_2$ in units of GeV$^{-2}$ extracted from the QCD sum rules 
in Eq.~(\protect\ref{we6}) (structure $\mbox{WE}_6$).} 
\label{g2uds}
\begin{tabular}{ccccccccc}
\hline\hline
 & $G_{2}^{u}$  & $G_{2}^{d}$ & $G_{2}^{s}$   & $G_{2}^{tot}$ & \\
\hline
$p \gamma \to \Delta^{+}$ & -1.57(32) & -0.77(10) & 0 & -2.34(48)\\
$n \gamma \to \Delta^{0}$  & 0.77(10) & 1.57(32) & 0 & 2.34(48)\\
$\Sigma^{+} \gamma \to \Sigma^{*+}$  & -0.05(37) & 0 & -0.36(10) & -0.41(45)\\
$\Sigma^{0} \gamma \to \Sigma^{*0}$ & -0.02(17) & 0.01(8) & -0.32(9) & -0.31(16)\\
$\Sigma^{-} \gamma \to \Sigma^{*-}$ & 0 & 0.11(20) & -0.43(10) & -0.32(12)\\
$\Xi^0 \gamma \to \Xi^{*0}$ & 0.15(24) & 0 & 0.23(6) & 0.38(29)\\
$\Xi^- \gamma \to \Xi^{*-}$ & 0 & -0.08(12) & 0.26(6) & 0.18(7)\\
$\Lambda \gamma \to \Sigma^{*0}$  & 0.42(11) & 0.20(6) & 0 & 0.62(17)\\
\hline\hline \end{tabular}
\end{table}

\subsection{Comparison with other theoretical calculations}

The determination of N to $\Delta$ transition amplitudes has been made 
in a number of other theoretical approaches.
In the following we briefly discuss each of these approaches 
in order to put our results from QCD sum rules in context.
The discussion is by no means exhaustive, but indicative of the breadth 
of interest in these transition amplitudes.
First, we present our results in Table~\ref{comp} after converting from $G_{1}$ 
from structure $\mbox{WE}_1$ and $G_{2}$ from $\mbox{WE}_6$ 
into the various conventions discussed earlier, and compare them with 
those from a lattice QCD calculation~\cite{Derek93}, 
a quark model calculation~\cite{Darewych83}, and experiment~\cite{PDG08}. 
For $G_{M1}$, we see that our results are slightly higher than those from lattice QCD 
and quark model calculations, 
but the overall pattern (in terms of magnitude and sign) is consistent among the three 
very different calculations.
For the ratio $R_{EM}$, the results are very different. We predict negative values 
for all transitions at the level of 10\% uncertainty. 
The lattice QCD results have too large errors to resolve the sign, although a 
more recent calculation in the $p\gamma \to \Delta^+$ channel gives 
a negative value~\cite{alex08}.  The quark model results have varying signs.
It demonstrates the difficulty of quantifying the small deformation from spherical 
symmetry in these transitions.

%
\begin{table*}[t] 
\caption{A comparison of N to $\Delta$ transition amplitudes from 
various approaches: this work (QCDSR) in three different conventions; 
lattice QCD~\cite{Derek93}; quark model~\cite{Darewych83}, and 
experiment~\cite{PDG08}.} \label{comp}
\begin{tabular}{c|ccccc|cc|cc|cc}
\hline\hline
Transition  &
\multicolumn{5}{c|}{QCDSR}&\multicolumn{2}{c|}{Lattice QCD}&\multicolumn{2}{c|}{Quark Model}
 &\multicolumn{2}{c}{Experiment} \\
 & $G_{M1}$ & $R_{EM}$ & $f_{M1}$ & $A_{1/2}$ & $A_{3/2}$ & $G_{M1}$ &
$R_{EM}$ & $G_{M1}$ & $R_{EM}$ & $G_{M1}$ & $R_{EM}$\\
 & $\mu_B$  & \% & GeV$ ^{-1/2}$ & GeV$ ^{-1/2}$ & GeV$ ^{-1/2}$ & $\mu_B$
 & \% & $\mu_B$  & \% & $\mu_B$  & \%\\
\hline
$p\gamma \to \Delta^+$ & 4.30(43)  & -1.66(17)  & 0.405  & -0.19  & -0.36
 & 2.46(43) & 3(8)  & 2.15 & -0.009  & 3.02 & -2.5(5)\\
$n\gamma \to  \Delta^0$ & -4.30(43)  & -1.66 (17) & -0.405  & 0.19  & 0.36
 & -2.46(43) & 3(8) & -2.15 & -0.009  &  & \\
$\Sigma^+ \gamma \to  \Sigma^{*+}$ & 4.18 (42) & -2.85(29)  & 0.240  & -0.11
 & -0.22  & 2.61(35) & 5(6)  & 2.61 & -0.210  &  & \\
$\Sigma^0 \gamma \to  \Sigma^{*0}$ & 1.79 (18) & -2.30 (23) & 0.104  & -0.05
 & -0.09  & 1.07(13) & 4(6)  & 1.1 & 0.192  &  & \\
$\Sigma^- \gamma \to  \Sigma^{*-}$ & -0.53 (5) & -7.99 (8) & -0.031  & 0.01
 & 0.03  & -0.47(9) & 8(4)  & -0.4 & 0.985  &  & \\
$\Xi^0 \gamma \to \Xi^{*0}$ & -1.69(17)  & -1.59(16)  & -0.094  & 0.04  &
0.08  & -2.77(31) & 2.4(2.7)  & -2.86 & 0.031  &  & \\
$\Xi^- \gamma\to  \Xi^{*-}$ & 0.19 (2) & -12.42 (13) & 0.010  & 0.00  &
-0.01  & 0.47(8) & 7.4(3.0)  & 0.44 & -0.259  &  & \\
$\Lambda \gamma \to \Sigma^{*0}$ & 3.34(34) & -4.62(48) & 0.314
&-0.135 &-0.285 & & &&
&&\\
\hline\hline
\end{tabular}
\end{table*}

\begin{figure}
\centerline{\vspace*{0.0cm}\psfig{file=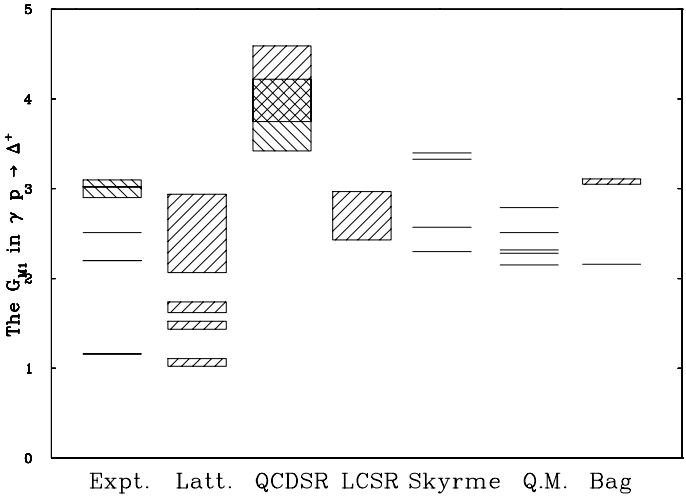,angle=90,width=9.0cm}}
\vspace*{0.0cm} \caption{A comparison of the magnetic dipole 
transition amplitude $G_{M1}$ in units of nucleon magnetons 
in the $p \gamma \to \Delta^{+}$ transition
from different approaches. See text for discussions.}
\label{comp-m1}
\end{figure}

\begin{figure}
\centerline{\vspace*{0.0cm}\psfig{file=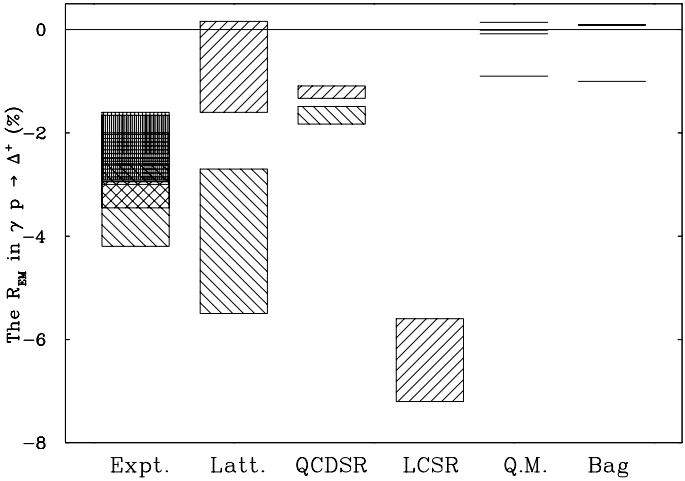,angle=90,width=9.0cm}}
\vspace*{0.0cm} \caption{A comparison of the $R_{EM}$ ratio in the $p \gamma \to \Delta^{+}$ transition. See text for discussions.}
\label{comp-rem}
\end{figure}

Since $p \gamma \to \Delta^{+}$ is the channel most widely studied and the 
only one measured in experiments, we give a more detailed comparison including 
more theoretical determinations.
Fig.~\protect\ref{comp-m1} summarizes the calculations of
the magnetic dipole transition amplitude $G_{M1}$ in units of nucleon magneton.
The calculations are grouped into six catagories: lattice QCD (Latt.),
QCD sum rules results from this work (QCDSR), light-cone QCD sum rules (LCSR), 
hedgehog models including Skyrme and hybrid models (Skyrme), 
quark model calculations (Q.M), 
and bag models (Bag).
The experimental result from the PDG is also displayed (Expt.).
Fig.~\protect\ref{comp-rem} summarizes the calculations of
the ratio $R_{EM}$ in the same channel.


There are only a few lattice QCD calculations. The original one by 
Leinweber {\it et. al.}~\cite{Derek93} obtained good signals for $G_{M1}$, 
but barely a signal for $R_{EM}$ ($+3\pm 8$ \%), and no signal for $G_{C2}$.
Shown here are the more recent calculations by Alexandrou {\it et. al.} 
which obtained reasonable signals for all three form factors.
For $G_{M1}$, from top down, in units of nucleon magneton ($\mu_N$), 
the results are $2.5\pm 0.4$~\cite{Derek93} with a quenched Wilson action; 
$1.68\pm 0.06$ with a hybrid action, $1.48\pm 0.04$ with a quenched Wilson action, 
and $1.07\pm 0.04$ with a 2-flavor Wilson action~\cite{alex08}. 
Strictly speaking, we should compare results at the chiral limit and $Q^2=0$.
The lattice results quoted are at pion mass of around 400 MeV and 
non-zero momentum transfer.
For the ratio $R_{EM}$, 
the results are $-4.1\pm 1.4$\%~\cite{alex04} at $Q^2=0.13$ GeV$^2$ 
extrapolated to the chiral limit in quenched QCD;
and $-0.77\pm 0.88$\%~\cite{alex08} at $Q^2=0.04$ GeV$^2$ and 
$m_\pi=360$ MeV in full QCD with a hybrid action.
The lattice signals for the transition form factors are by now well establsihed.
With increasing computing power and better lattice technology, 
lattice calculations offer the best promise for precision determination of 
the transitions from QCD.

The two numbers for QCDSR are from different sum rules: 
the top one correponds to $G_1$ from WE$_1$ and $G_2$ from 
WE$_6$, the bottom one correponds to $G_1$ from WO$_2$ and $G_2$ from WE$_6$.
They are converted to $G_{M1}$ and $R_{EM}$ for comparison.

The ligh-cone QCD results come from Ref.~\cite{Rohrwild07} which involve 
photon distribution amplitudes (light-cone wavefunctions) up to twist 4.
  
The hedgehog models from top down include the 
holographic QCD calculation by Grigoryan, Lee and Yee~\cite{Grigoryan09} which 
can be considered as a 5D version of the Skyrme model;
the SU(3) Skyrme model calculation by Chemtob~\cite{Chemtob85};
the SU(2) Skyrme model calculations by Kunz and Mulders ~\cite{Kunz90}; 
and the SU(2) Skyrme model calculation by Adkins, Nappi and Witten~\cite{Adkins83}. 
The semiclassical hedgehog calculation of Cohen and Broniowski~\cite{Cohen86} 
gives a negative value (-2.8) for M1 moment so it is not shown in Fig.~\ref{comp-m1}.
Skyrme models produce reasonable results for the M1 amplitude, but 
vanishing ones for eletric quadrupole E2 transition matrix elements 
so they are left out in Fig.~\ref{comp-rem}.

The quark model results include (from top down): a calculation by 
Franklin~\cite{Franklin02} (It has been correlated with $G_A$ and provides 
arguably the best constraint on $G_{M1}$ in quark-model calculations);
a Bethe-Salpeter determination from Mitra and Mittal~\cite{Mitra}; 
a calculation by Guiasu and Koniuk~\cite{Guiasu87} in which mesonic dressings 
of the nucleon are explicitly included; a calculation by Capstick ~\cite{Capstick92} in
which configuration mixing in the baryon SU(6) wave functions is
accounted for; and a calculation by Darewych {\it et. al.}~\cite{Darewych83} 
based on the simple quark model.
Quark model calculations generally give small values for $R_{EM}$.

The bag models include a chiral bag calculation (top)  
by Kalbermann and Eisenberg~\cite{Kalber}, and 
an old MIT bag calculation (bottom) by Donoghue {\it et. al.}~\cite{Donoghue}. 

\section{Conclusion}
\label{con}

We have carried out a comprehensive study of the $N\gamma\to\Delta$ transition
amplitudes using the method of QCD sum rules.
We derived a new, complete set of QCD sum rules using generalized interpolating fields
and examined them by a Monte-Carlo analysis.
Here is a summary of our findings.

We proposed a new way of extracting the transition amplitudes from the
slope of straight lines as a function of $1/M^2$ 
in conjunction with the corresponding mass sum rules, 
as defined in Eq.~(\ref{ratio}). 
We find that this method  is more robust than from
looking for `flatness' as a function of Borel mass. 
The linearity displayed from the OPE side
matches well with the phenomenological side in most
cases. The method also demonstrates clearly the role of the non-diagonal transition
terms in the intermediate states caused by the external field: 
wherever such transitions are large, the corresponding sum rules perform poorly.

Of the 96 sum rules we derived (from 12 independent structures, each for 8 transitions), 
we find that the sum rules from
the $\mbox{WE}_1$ and $\mbox{WO}_2$ structures are the most reliable for the transition 
amplitude $G_1$, based on OPE convergence and ground-state pole dominance, and smallness 
of the non-diagonal transitions.
The QCD sum rules from these structures are in Eq.~(\ref{we1}) and Eq.~(\ref{wo2});
their predictions are found in Table~\ref{tabwe1} and Table~\ref{tabwo2}.
Our attempt to extract the $G_2$ amplitude was less successful. The only sum rules 
that can give stable results are in Eq.~(\ref{we6}) from structure $\mbox{WE}_6$.
Their predictions are found in Table~\ref{tabwe6}.
Our final results in the various conventions are found in Table~\ref{comp}.

Our Monte-Carlo analysis revealed that there is an uncertainty
on the level of 10\% in the transition amplitudes if we assign 10\% uncertainty
in the QCD input parameters. It goes up to about 30\% if we adopt the
conservative assignments that have a wide range of uncertainties in Ref.~\cite{Derek96}.
The Monte-Carlo analysis also revealed some correlations between the input and output
parameters. The most sensitive is the vacuum susceptibility $\chi$. So a better determination
of this parameter can help improve the accuracy on the transition amplitudes and other
quantities computed from the same method.
We also isolated the individual quark contributions to the
transition amplitudes. These contributions provide insight into 
the effects of SU(3)-flavor symmetry breakings in the strange quark,
and environment sensitivity of quarks in different baryons.

We compared our results with a variety of theoretical calculations, and with 
experiment in the proton channel. Our result for transition amplitude $G_{M1}$ 
is larger than the experiment and other calculations, 
while our result for the ratio $R_{EM}$ is consistent 
with experiment. In general, we find that $G_1$ amplitudes are more stable 
than $G_2$ amplitudes in this approach. 
Our results for the $\Lambda \gamma \to \Sigma^{*0}$ transition are new.

Overall, the results for the N to $\Delta$ transition amplitudes in the QCD sum rule 
approach are not as robust as those for the baryon magnetic 
moments~\cite{Wang08,Lee98b,Lee98c}. 
It is in large part due to the intrinsic un-equal mass 
double pole in the phenomenological representation of the spectral functions. 
Nonetheless, the calculations offer a physically-transparent, QCD-based perspective 
on the transitions in terms of quarks, gluons and vacuum condensates.

\begin{acknowledgments}
This work is supported in part by U.S. Department of Energy under grant
DE-FG02-95ER-40907.

\end{acknowledgments}


\end{document}